\documentclass[aps,prx,twocolumn,a4paper,showkeys,10pt,longbibliography]{revtex4-1}
\usepackage[english]{babel}
\usepackage[utf8x]{inputenc}
\usepackage[T1]{fontenc}

\usepackage[a4paper,top=3cm,bottom=2cm,left=2cm,right=2cm,marginparwidth=1.75cm]{geometry}

\usepackage{physics}
\usepackage{float}
\usepackage{etoolbox}
\usepackage{graphicx}
\graphicspath{{\images}}
\usepackage{subfigure}
\usepackage{tikz}
\usepackage{hyperref}
\apptocmd{\sloppy}{\hbadness 10000\relax}{}{} 

\usepackage{amsmath}
\usepackage{amsfonts}
\usepackage{amssymb}
\usepackage{amsthm}
\usepackage{mathtools}
\usepackage{pgfplotstable}
\usetikzlibrary{arrows,arrows.meta,shapes,snakes,automata,backgrounds,petri}

\DeclareMathOperator*{\argmin}{arg\,min}

\theoremstyle{definition}

\theoremstyle{definition}

\newcommand{\XUV}{\mathrm{XUV}}
\newcommand{\IR}{\mathrm{IR}}

\begin{document}
\title{Propensity rules and interference effects in laser-assisted photoionization of helium and neon}
\author{Mattias \surname{Bertolino}}
\affiliation{Department of Physics, Lund University, Box 118, SE-221 00 Lund, Sweden}
\author{David \surname{Busto}}
\affiliation{Department of Physics, Lund University, Box 118, SE-221 00 Lund, Sweden}
\author{Felipe \surname{Zapata}}
\affiliation{Department of Physics, Lund University, Box 118, SE-221 00 Lund, Sweden}
\author{Jan~Marcus \surname{Dahlstr\"om}}
\affiliation{Department of Physics, Lund University, Box 118, SE-221 00 Lund, Sweden}
\date{\today}

\begin{abstract}
We investigate the angle-resolved photoelectron spectra from laser-assisted photoionization for helium and neon atoms using an \textit{ab initio} method based on time-dependent surface flux and configuration interaction singles. We find that the shape of the distributions can be interpreted using a propensity rule, an intrinsic difference in the absorption and emission processes, as well as interference effects between multiple paths to the final angular momentum state.
In neon we find that the difference between absorption and emission is hidden in the first sideband due to the multiple competing $m$ channels. Together, this aids the understanding of the formation of minima in the angular distributions, which can be transferred to an improved understanding on photoionization time delays in attosecond science.
\end{abstract}

\maketitle

\section{Introduction}
Photoionization is a fundamental process in nature  where absorption of a photon by an atom leads to the emission of a photoelectron and to the creation of a positive ion: $A+\gamma \to A^+ +e^-$. The photoelectric effect is possible only when the energy of the photon, $\hslash \omega$, is sufficiently large to overcome the electron binding energy of the atom, $I_\mathrm{p}$. Although the process is explained qualitatively as a one-electron transition from an occupied atomic orbital to a continuum of states, the quantitative photoionization rates are affected by electron-electron correlation effects in the atom, as evidenced in works based on many-body perturbation theory ~\cite{starace_theory_1982, Amusia1990,KheifetsPRA2013}. 
Given photons with a wavelength that is larger than the size of the atom, $a_0=5.29\times10^{-11}$\,m, the electronic transitions can be simplified to follow the dipole {\it selection rules}, where the angular momentum of the electron must change by one unit: $\Delta\ell = \pm 1$. This reduces the complexity of the problem to a finite set of continua that can be reached by the photoelectron. As a complement to the dipole selection rules, Fano proposed a {\it propensity rule}, which states that absorption of light favours increasing the electron angular momentum, $\ell \to \ell+1$, over decreasing the angular momentum, $\ell \to \ell-1$ ~\cite{fano_propensity_1985}. This simple rule explains why in neon the probability to reach the $d$-wave from the initial $2p$ orbital is larger than the probability to reach the $s$-wave. As with most simple rules, there are also some notable  exceptions, e.g.\ in photoionization from the $3p$ orbital in argon in the vicinity of the Cooper minimum, where the $d$-wave contribution becomes very small due to a vanishing dipole element for the transition~\cite{cooper:1962}.  

In recent years, novel light sources have made it possible to study light--matter interactions in more extreme conditions where the atoms are subject to more intense short-wavelength fields~\cite{Seddon_2017}, multi-color fields~\cite{Allaria2013,LutmanPRL2013,GauthierPRL2016} and short pulses with duration on the femtosecond and attosecond time scale~\cite{KrauszRMP2009}. 
One class of problems that has attracted attention is {\it laser-assisted photoionization}, where an atom is photoionized using radiation of short wavelength, typically extreme ultraviolet radiation (XUV), but with an additional long-wavelength laser field, typically in the infrared range (IR), which \textit{dresses} the atom: $A + \gamma_\XUV \pm q \gamma_\IR \to A^+ + e^-$. 
In this case, the electron is ionized by the XUV field and then subsequently interacts with the IR field leading to laser-driven continuum--continuum transitions. 
In the multi-cycle pulse limit, the resulting photoelectron spectrum includes a main peak at an energy given by the XUV photoelectric effect and a number of sideband peaks due to the increasing number of interactions with the IR field: $E_{\mathrm{kin}} = \hslash \omega_{\mathrm{XUV}} \pm q \hslash \omega_{\mathrm{IR}} - I_\mathrm{p}$. 
In the case where the IR field is weak, the strength of the sidebands decreases with each order as expected from perturbation theory with a probability determined by the intensity power law in atomic units: $P_q \propto (I_\IR)^{q}$. 
In the opposite case, when the IR field is strong, laser-assisted photoionization can be interpreted using semi-classical electron trajectories~\cite{Dusterer_interference_2013, Dusterer_two-color_2019}. 

Laser-assisted photoionization has been studied analytically using time-dependent Volkov states, which by their closed-form solution allow for efficient calculations of cross-sections for laser-assisted scattering and ionization~\cite{kroll_charged-particle_1973, MadsenAJP2005,AlvaroNJP2013}. 
More accurate numerical studies have been performed by perturbation theory within the single-active electron (SAE) approximation \cite{TomaJPB2002,DahlstromCP2013} and by many-body perturbation theory at the level of one-photon Random Phase Approximation with Exchange (RPAE) with uncorrelated continuum--continuum transitions for closed shell atoms \cite{DahlstromPRA2012} and for photodetachment of negative ions \cite{LindrothPRA2017}. 
Recently, a gauge-invariant two-photon RPAE approach has been demonstrated \cite{VinbladhPRA2019}. 
Numerical simulations have also been performed in the time domain within the SAE \cite{ZhangPRA2010,NagelePRA2011,IvanovPRA2017,BrayPRA2018}, for helium \cite{PazourekPRL2012} and many-electron atoms, e.g.\ neon by R-matrix theory  ~\cite{MoorePRA2011, Lysaght_time-dependent_2009} and argon by Density Functional Theory (DFT) \cite{Sato2018}. 
Many-electron correlations in non-linear photoionization has also been studied by time-dependent theories based on Configuration Interaction Singles (CIS) ~\cite{Karamatskou_2014} and Multi-Configuration Self-Consistent Fields (MCSCF)~\cite{orimo_application_2019}.  

Laser-assisted photoionization is an important process in attosecond science, where it is at the core of both pulse characterization techniques using the RABBIT technique~\cite{PaulScience2001} and for measurement of atomic delays in photoionization~\cite{SchultzeScience2010,isinger_photoionization_2017}. 
Recently, atomic delay measurements have been performed with angular resolution~\cite{HeuserPRA2016,busto_fano_2019,cirelli_anisotropic_2018}. 
This has evidenced that subtle differences in absorption and emission processes in the continuum--continuum transitions can lead to a strong dependency on the atomic delay with angle of emission, incomplete quantum interference in RABBIT measurements and to qualitatively different angular distributions of photoelectrons, as explained by Busto et al.\ by extending Fano's propensity rule to continuum--continuum transitions~\cite{busto_fano_2019}. 

In this paper, we perform {\it ab-initio} simulations of laser-assisted photoionization by propagating the Time-Dependent Schrödinger Equation (TDSE) within the Configuration Interaction Singles approximation (TDCIS)~\cite{RohringerPRA2006, GreenmanPRA2010}. 
This allows us to examine the angular distributions and propensity rules in laser-assisted photoionization of helium and neon atoms for both the first sideband and higher-order sidebands generated by absorption of multiple IR photons. 
We find different angular distributions formed by absorption and emission processes in the continuum, and we are able to verify that the propensity rules can be extended to higher-order continuum--continuum transitions driven by the IR field. 
The paper is structured as follows.
In Section~\ref{sec:method}, our method is described along the relevant laser parameters.
In Section~\ref{sec:results}, we present our results of the numerical simulations.
Finally, in Section~\ref{sec:conclusions}, we draw conclusions of the presented data and discuss potential topics for future studies.
Atomic units ($\hslash = e = m_e = 4\pi\varepsilon_0=1$) are used through out this paper if not specifically stated.

\section{Method}
\label{sec:method}
In this section we describe our method to compute laser-assisted photoionization from closed-shell atoms. In part A we describe the vector potential used to model the electromagnetic fields, in part B we review the TDCIS ansatz, in part C we present details of our t-SURFF implementation and in part D we give some more details on our numerical implementation.   
\subsection{Field description}
\label{subsec:method-field-description}
The numerical experiments are carried out with Gaussian XUV- and IR-pulses, linearly polarized along the quantization axis $\hat{z}$, that are overlapped in time and defined by a vector potential given by
\begin{align}
    A =& \left[A_0^{\mathrm{XUV}} \sin(\omega_{\mathrm{XUV}}t) + A_0^{\mathrm{IR}} \sin(\omega_{\mathrm{IR}}t)\right] \nonumber \\ \times& \exp\left[-2\ln(2) \frac{t^2}{\tau^2}\right],
\end{align}
where $A_0^{\mathrm{XUV}} = 0.005$\,a.u.\ and $A_0^{\IR} = 0.003$\,a.u.\ which yields an peak intensity of the IR pulse of $5.6\times10^9$\,W/cm$^2$. 
This intensity implies only perturbative action by the IR-field.
The duration of the pulses is given by $\tau = 410 \text{\,a.u}.\,\approx 10$\,fs and the frequency of the IR-field is given by $\omega_{\IR} \approx 1.55$\,eV to match a Ti-Saph.\ laser system.
The fact that the XUV pulse duration is longer than the IR period,  $\tau > 2\pi/\omega_\mathrm{\IR}$, implies that the photoelectron spectrum will consist of discrete peaks that correspond to interaction with $q$ photons in the continuum. 

\subsection{TDCIS ansatz}
\label{subsec:method-tdcis}
The TDCIS ansatz~\cite{RohringerPRA2006} for the many electron wave function is
\begin{equation}
    \label{eq:cis}
    \ket{\Psi(t)} = \alpha_0(t) \ket{\Phi_0} + \sum_a^{\mathrm{occ}} \sum_p^{\mathrm{exc}} \alpha_a^p(t) \ket{\Phi_a^p},
\end{equation}
where $\ket{\Phi_0}$ is the Hartree-Fock ground state and the singly excited states $\ket{\Phi_a^p}$ are constructed using the framework of second quantization, 
\begin{equation}
    \ket{\Phi_a^p} = \frac{1}{\sqrt{2}} \{ \hat{c}_{p+}^\dagger \hat{c}_{a+} + \hat{c}_{p-}^\dagger \hat{c}_{a-} \}\ket{\Phi_0},
\label{eq:Phipa}
\end{equation}
where $\hat{c}_{p\sigma}^\dagger$ creates an electron in the virtual (exc) orbital $p$ with spin $\sigma$, $\ket{\phi_{p\sigma}}$, while $\hat{c}_{a\sigma}$ creates a hole in the initially occupied (occ) orbital $a$ with spin $\sigma$, $\ket{\varphi_{a \sigma}}$.
The ansatz in Eq.~\eqref{eq:Phipa} assures that the spin {\it singlet} state character of the closed-shell initial state is maintained also for excited states. Similarly, we adopt the {\it gerade} formulation of TDCIS to make full use of the symmetry in magnetic quantum numbers, $m_p=m_a$, for linearly polarized fields \cite{PabstPRA2012}.  
In TDCIS the time dependence is found only in the complex amplitudes, $\alpha_0(t)$ and $\alpha^p_a(t)$, and the static orbitals are found by solving the mean-field HF problem without fields present. The time evolution of the complex amplitudes is found by projecting the ansatz in Eq.~\eqref{eq:cis} onto the TDSE with a laser-interaction Hamiltonian, as shown in Ref.~\cite{GreenmanPRA2010,PabstPRA2012}.
Here, we consider light-matter interaction within the dipole approximation given by $V_I(t)=A_z(t) \hat p_z$, where $E_z(t)=-\partial A_z(t)/\partial t$ is the electric field with linear polarization along the $\hat z$-direction.  
The choice of helium and neon is done on the basis of their different initial angular momentum state and hence their difference in the accessible continuum-state after the absorption of one XUV-photon. In addition, both helium and neon are well-described by the truncated basis of the TDCIS theory.

\subsection{Implementation of t-SURFF with TDCIS}
\label{subsec:implementation-tsurff-tdcis}
Within TDCIS theory, the excited many-body state can be expressed as one-electron time-dependent orbitals~\cite{RohringerPRA2006},
\begin{equation}
    \chi_{a}(\mathbf{r},t) = \sum_{p}^\mathrm{exc}\alpha^p_a(t)\phi_p(\mathbf{r}),
\end{equation}
associated with each created hole, $a$.
The time-dependent orbitals can also be expanded as
\begin{equation}
    \label{eq:td-orbitals-expansion}
    \chi_{a}(\mathbf{r},t) = \frac{1}{r} \sum_{\ell_p} \psi_{\ell_p m_a}^a (r,t) Y_{\ell_p m_a}(\Omega_{\mathbf{r}}), 
\end{equation}
where $\ell_p$ runs over all possible angular momenta attainable by the electron.
The t-SURFF method relies on knowledge of the photoelectron wavefunction, $\chi_{a}(\mathbf{r}_c,t)$, at a given radius, $r_{c}<r_{\mathrm{ecs}}$, at all times, $t$, and then makes use of the approximate Volkov states to account for field-induced dynamics of the photoelectron beyond $r_c$ \cite{TaoNJP2012}. 
We use a modified Volkov Hamiltonian, 
\begin{equation}
    \label{eq:volkov-hamiltonian}
    \hat{H}^{(V)}_a(t) = \frac{\hat p^2}{2} +{A_z}(t)\hat p_z - \varepsilon_a,
\end{equation}
where $\hat p^2=-\nabla^2$ and $\hat p_z=-i\partial/\partial z$, to model the dynamics of the time-dependent orbital in the region far from the ion, where Coulomb interactions can be neglected.
The energy of the photoelectron depends on the binding energy of orbital $a$ in accordance with Koopman's theorem, $I_\mathrm{p}=-\varepsilon_a$.
The time-dependent orbitals that satisfy the TDSE with the  Hamiltonian from Eq.~\eqref{eq:volkov-hamiltonian} are
\begin{align}
  \label{eq:volkov-state}
\chi^{(V)}_{\mathbf{k},a}(\mathbf{r},t) = \frac{1}{(2\pi)^{3/2}}
\exp[i\mathbf{k}\cdot\mathbf{r}] \nonumber \\
\times \exp\left[ -i\int_{t_\mathrm{ref}}^t \dd t'\left\{ \frac{k^2}{2} +{A_z}(t'){k_z}-\varepsilon_a\right\}\right], 
\end{align}
which are plane waves with a time-dependent phase.

The spectral amplitudes for laser-assisted photoionization are found using a complex amplitude for the overlap of time-dependent orbitals in the outer region, $r>r_c$, using a radial Heaviside operator acting at $r_c$, $\hat{\theta}(r_c)$, defined as   
\begin{equation}
\label{eq:bka}
b_{\mathbf{k},a}(t) = \mel{\chi^{(V)}_{\mathbf{k},a}(t)}{\hat{\theta}(r_c)}{\chi_{a}(t)}.
\end{equation}
The complex amplitude in Eq.~\eqref{eq:bka} becomes the scattering amplitude when the time 
is evaluated at a late hour, $t=T$, after which all external fields have ended and after which the photoelectron wave packet have propagated far away from the ion \cite{TaoNJP2012}. 
We obtain a final expression for the scattering coefficients in t-SURFF given by
\begin{widetext}
\begin{align}
  \label{eq:TSURFF-final}
  \begin{split}
  b_{\mathbf{k},a}(T)
  &= i ~\sqrt[]{\frac{2}{\pi}} \int_{-\infty}^T \dd t~ \exp\left[i \int_{t_{\mathrm{ref}}}^t \dd \tau \left\{ \frac{k^2}{2} + A_z(\tau)k_z - \varepsilon_a \right\} \right] \\
  & \times \sum_{\ell_p} \bigg{\{} (-i)^{\ell_p} \frac{1}{2} \left(kr_c j_{\ell_p}^\prime(kr_c) + 
    j_{\ell_p}(kr_c) \right) \psi_{\ell_p m_a}^a(r_c,t) Y_{\ell_p m_a}(\Omega_{\mathbf{k}}) \\
  &- \frac{(-i)^{\ell_p}}{2} r_c j_{\ell_p}(kr_c) \psi_{\ell_p m_a}^{a \prime} (r_c,t) Y_{\ell_p m_a}(\Omega_{\mathbf{k}}) \\
  & + \frac{i}{2~\sqrt[]{\pi}} r_c A_z(t) ~\sqrt[]{2\ell_p+1} \psi_{\ell_p m_a}^a(r_c,t)
  \sum_{\ell=\ell_p \pm 1} (-i)^\ell \frac{j_{\ell}(kr_c)}{2\ell+1} C_{\ell_p m_a,10}^{\ell m_a} C_{\ell_p 0,10}^{\ell 0} 
Y_{\ell m_a}(\Omega_{\mathbf{k}}) \bigg{\}},
  \end{split}
\end{align}
\end{widetext}
where $j_{\ell}(kr)$ is the spherical bessel function of order $\ell$ and $j_{\ell}^{\prime}(kr)$ is its derivative, both evaluated at $kr$.
Further $Y_{\ell,m}(\Omega_{\mathbf{k}})$ is the spherical harmonic of order $\ell$ and $m$, evaluated at angle $\Omega_{\mathbf{k}} \equiv (\theta_{\mathbf{k}}, \varphi_{\mathbf{k}})$, and $C_{\ell m, 10}^{\ell^\prime m}$ is the Clebsch-Gordan coefficient for a dipole transition with linear polarized light.
We note that the t-SURFF method is an approximate method for analysis of photoelectrons when applied to this problem with a long-range potential from the remaining ion~\cite{TaoNJP2012}. Single or multiphoton transitions that populate Rydberg states can also be expected to cause some problems in special cases due to their large radial extent, $\sim n^2$. 

The angular distribution of the photoelectron can be described by a coherent superposition of partial waves with the corresponding angular momenta determined by dipole selection rules. 
As an example, the case of two final angular momenta, which is expressed by two spherical harmonics: $Y_{\ell>,m}$ and $Y_{\ell<,m}$, with $\ell_>=\ell_0+1$ and $\ell_<=\ell_0-1$, has a complex amplitude with an angle-dependence given by
\begin{equation}
    \label{eq:fthetavarhpi}
    f(\theta) = 
    \tilde a_{\ell_>}Y_{\ell_>,m}(\theta,\varphi)+
    \tilde a_{\ell_<}Y_{\ell_<,m}(\theta,\varphi).
\end{equation}
To find these partial wave amplitudes from the general scattering amplitudes, $b_{\mathbf{k},a}(T)$, we solve a minimization problem,
\begin{equation}
    \label{eq:minimization}
    \tilde{a} = \argmin_{a} \sum_{i} \bigg{|}f_q(\theta_i) - \sum_{m}|\sum_{\ell} a_{\ell m} Y_{\ell m}(\theta_i, \varphi)|^2 \bigg{|}^2,
\end{equation}
for the general complex amplitudes $\tilde{a} = \{\tilde{a}_{\ell m}\}$. The magnetic quantum number of the photoelectron is linked to that of the hole, $m_p=m_a=m$, which is typically unresolved in experiments and, therefore, is summed over incoherently. In Eq.~\eqref{eq:minimization} the angular probability distribution of a given peak $q$ is computed by integrating over energy, that is
\begin{equation}
    \label{eq:peak-integration}
    f_q(\theta_i) = \frac{1}{2} \int_{E_q - \xi}^{E_q + \xi} \dd E ~ |b_{\mathbf{k}_i,a}(T)|^2,
\end{equation}
where $E_q$ is the energy at the center of the peak $q$, using Eq.~\eqref{eq:bka} for a given final momentum of the photoelectron, $k=|\mathbf{k}|$ evaluated at a set of polar angles $\theta_i$. This procedure allows us to extract partial wave amplitudes for all photoelectron peaks, $\pm q$, which we label by $\tilde{a}_{\ell m}^{\pm q}$, where the reference to $m$ is sometimes is omitted for brevity.  

\subsection{Numerical implementation}
\label{subsec:methods-numerical-implementation}
Our method is similar to that of Karamatskou et al.~\cite{Karamatskou_2014}, as it combines TDCIS for closed-shell atoms~\cite{GreenmanPRA2010} with the Time-Dependent Surface Flux (t-SURFF) method~\cite{TaoNJP2012}, but our method differs in a number of ways: (i) our numerical implementation is based on B-splines~\cite{deboor}, (ii) we use Exterior Complex Scaling (ECS) to handle the boundary conditions of the outgoing photoelectrons \cite{SimonPLA1979}, (iii) our implementation of t-SURFF, Eq.~\eqref{eq:TSURFF-final}, differs in its detailed derivation, as discussed in Appendix~\ref{apdx:tsurff}. 

In this work we restrict the active space configuration in energy, $E^p_a=\varepsilon_p-\varepsilon_a<30 \text{\,a.u.\,} = 816.33$\,eV, and use an electron angular momentum of at least $\ell = 6$.
We also restrict the TDCIS calculation to the outermost valence orbital in the sum over occ in Eq.~\eqref{eq:cis} and, therefore, do not consider XUV-stimulated hole--hole transitions that can lead to further excitation of the ion within TDCIS~\cite{YouPRA2016}.
The latter restriction implies that we consider the $1s$ orbital in helium, but only excitation from the $2p$ orbital in neon.
The binding energies used are the Hartree-Fock binding energies $I_p = 0.918$\,a.u.\ for helium and $I_p = 0.850$\,a.u.\ for the $2p$ orbital in neon.

For the B-spline interpolation, we use 165 and 320 knotpoints in the inner region for helium and neon respectively and 30 knotpoints in the ECS region for both atoms.
The polynomials used are chosen to be of order 6.
We use a knotpoint spacing of 0.4\,a.u.\ and an ECS-angle of 25 degrees.
The use of ECS leads to a non-Hermitian Hamiltonian, where the virtual states are exponentially damped in time by complex eigenvalues. In space the electron wavefunctions remain physical within the radius $r_{\mathrm{ecs}} = 64$\,a.u.\ for helium and $r_{\mathrm{ecs}} = 120$\,a.u.\ for neon. Inside the ECS region, $r>r_{\mathrm{ecs}}$, the photoelectron wavefunction is damped radially, which helps to remove nonphysical reflections from the end point of the radial knotpoint sequence. The use of ECS restricts the propagation of TDCIS to the velocity gauge.

\section{Results}
\label{sec:results}
In this section we present our results from laser-assisted photoionization simulations.
We present the numerically obtained photoelectron angular distributions (PAD) for helium in part~A and for neon in part~B.
The frequency of the XUV-photon, $\omega_{\XUV}$, is varied in order to study how the PAD depends on different final kinetic energies of the photoelectron. 

\begin{figure}
\begin{center}
\begin{tikzpicture}[scale=0.8, transform shape]
  \node at (3.5,5.5) {\large $(a)$};
  \node at (1.5,5.5) {\large He};
  \node (peaks) at (-1,5.5) {\large Peaks (q)};
  \node (p2) at (-1,5) {$2$};
  \node (p1) [below of=p2]     {$1$};
  \node (c)  [below of=p1]     {$0$};
  \node (m1) [below of=c] {$-1$};
  \node (m2) [below of=m1] {$-2$};
  
  \coordinate (O) at (0,0);
  \coordinate (a) at (1,3);
  \coordinate (aa) at (2,4);
  \coordinate (ab) at (0,4);
  \coordinate (aaa) at (3,5);
  \coordinate (aab) at (1,5);

  \draw[thick] (-1,0.5) -- (1,0.5);
  \foreach \x in {-1,-0.9,...,1} {
    \draw[-,help lines] (\x,0.5) -- (\x+0.1,0.6);
  }

  \draw (-0.5,0) -- (0.5,0);
  \node[below] (O) {$s$};
  \draw[->, color=blue] (O) -- (a) node[below, color=black] {$p$};
  \draw (0.5,3) -- (1.5,3);
  \draw[->, color=red] (a) -- (aa) node[below, color=black] {$d$};
  \draw (-0.5,4) -- (0.5,4);
  \draw[->, color=red, dashed] (a) -- (ab) node[below, color=black] {$s$};
  \draw (1.5,4) -- (2.5,4);
  \draw[->, color=red] (aa) -- (aaa) node[below, color=black] {$f$};
  \draw (0.5,5) -- (1.5,5);
  \draw[->, color=red] (ab) -- (aab) node[below, color=black] {$p$};
  \draw (2.5,5) -- (3.5,5);
  \draw[->, color=red, dashed] (aa) -- (aab);
  
  \coordinate (ee) at (2,2);
  \coordinate (eb) at (0,2);
  \coordinate (eee) at (3,1);
  \coordinate (eeb) at (1,1);
  
  \draw[->, color=red, dashed] (a) -- (ee) node[below, color=black] {$d$};
  \draw (-0.5,2) -- (0.5,2);
  \draw[->, color=red] (a) -- (eb) node[below, color=black] {$s$};
  \draw (1.5,2) -- (2.5,2);
  \draw[->, color=red, dashed] (ee) -- (eee) node[below, color=black] {$f$};
  \draw (0.5,1) -- (1.5,1);
  \draw[->, color=red, dashed] (eb) -- (eeb) node[below, color=black] {$p$};
  \draw (2.5,1) -- (3.5,1);
  \draw[->, color=red] (ee) -- (eeb);
\end{tikzpicture}
\begin{tikzpicture}[scale=0.8, transform shape]
  \node at (4.5,5.5) {\large $(b)$};
  \node at (2,5.5) {\large Ne};
  \node (peaks) at (-1,5.5) {\large Peaks (q)};
  \node (p2) at (-1,5) {$2$};
  \node (p1) [below of=p2]     {$1$};
  \node (c)  [below of=p1]     {$0$};
  \node (m1) [below of=c] {$-1$};
  \node (m2) [below of=m1] {$-2$};

  \coordinate (ON) at (1,0);
  \coordinate (0sN) at (0,3);
  \coordinate (0dN) at (2,3);
  \coordinate (1fN) at (3,4);
  \coordinate (1pN) at (1,4);
  \coordinate (2sN) at (0,5);
  \coordinate (2dN) at (2,5);
  \coordinate (2gN) at (4,5);

  \draw[thick] (0,0.5) -- (2,0.5);
  \foreach \x in {0,0.1,...,2} {
    \draw[-,help lines] (\x,0.5) -- (\x+0.1,0.6);
  } 

  \draw (0.5,0) -- (1.5,0);
  \node at (1,-0.2) {$p$};
  \draw[->, color=blue, dashed] (1,0) -- (0sN) node[below, color=black] {$s$};
  \draw[dotted, thick] (-0.5,3) -- (0.5,3);
  \draw[->, color=blue] (1,0) -- (0dN) node[below, color=black] {$d$};
  \draw (1.5,3) -- (2.5,3);
  \draw[->, color=red] (0sN) -- (1pN);
  \draw[->, color=red] (0dN) -- (1fN) node[below, color=black] {$f$};
  \draw (0.5,4) -- (1.5,4);
  \draw[->, color=red, dashed] (0dN) -- (1pN) node[below, color=black] {$p$};
  \draw (2.5,4) -- (3.5,4);
  \draw[->, color=red, dashed] (1pN) -- (2sN) node[below, color=black] {$s$};
  \draw[dotted, thick] (-0.5,5) -- (0.5,5);
  \draw[->, color=red] (1pN) -- (2dN) node[below, color=black] {$d$};
  \draw[->, color=red, dashed] (1fN) -- (2dN);
  \draw (1.5,5) -- (2.5,5);
  \draw[->, color=red] (1fN) -- (2gN) node[below, color=black] {$g$};
  \draw (3.5,5) -- (4.5,5);

  \coordinate (m1pN) at (1,2);
  \coordinate (m1fN) at (3,2);

  \draw[->, color=red, dashed] (0sN) -- (m1pN);
  \draw[->, color=red] (0dN) -- (m1pN) node[below, color=black] {$p$};
  \draw (0.5,2) -- (1.5,2);
  \draw[->, color=red, dashed] (0dN) -- (m1fN) node[below, color=black] {$f$};
  \draw (2.5,2) -- (3.5,2);
  \draw[->, color=red, dashed] (m1fN) -- (4,1) node[below, color=black] {$g$};
  \draw (1.5,1) -- (2.5,1);
  \draw[->, color=red, dashed] (m1pN) -- (2,1) node[below, color=black] {$d$};
  \draw (3.5,1) -- (4.5,1);
  \draw[->, color=red] (m1fN) -- (2,1);
  \draw[->, color=red] (m1pN) -- (0,1) node[below, color=black] {$s$};
  \draw[dotted, thick] (-0.5,1) -- (0.5,1);

\end{tikzpicture}
\end{center}
\caption{Laser-assisted photoionization paths in (a) helium and (b) neon in the case of absorption or emission of IR-photons. In neon, the dotted $s$-states indicate that they are only reachable for the case of $m=0$. The propensity rule is illustrated with filled lines as these transitions are relatively more probable than the transitions of dashed lines when comparing absorption and emission.}
\label{fig:sketch}
\end{figure}
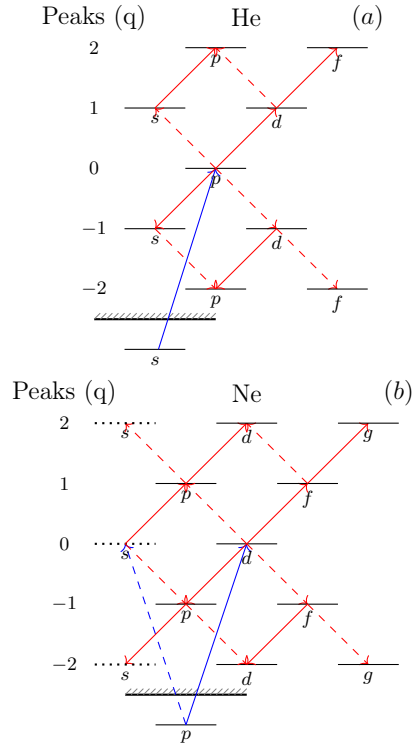

In Fig.~\ref{fig:sketch}, the laser-assisted photoionization paths are shown for helium and neon respectively.
The main peak in the photoelectron spectrum is originating from absorption of one XUV photon and it is denoted $q=0$. 
The sidebands corresponding to additional absorption $(+)$ and emission $(-)$ of $q$ IR photons are denoted by $\pm q$.  
The photoelectron alters its orbital angular momentum by plus or minus one for each interaction event with the dressing IR-field. 
The PAD results from different spherical harmonics in superposition, as shown for each value of $q$ in Fig~\ref{fig:sketch}.

\subsection{Helium}
\label{subsec:results-helium}
\begin{figure*}
  \centering
  {\includegraphics[scale=0.29]{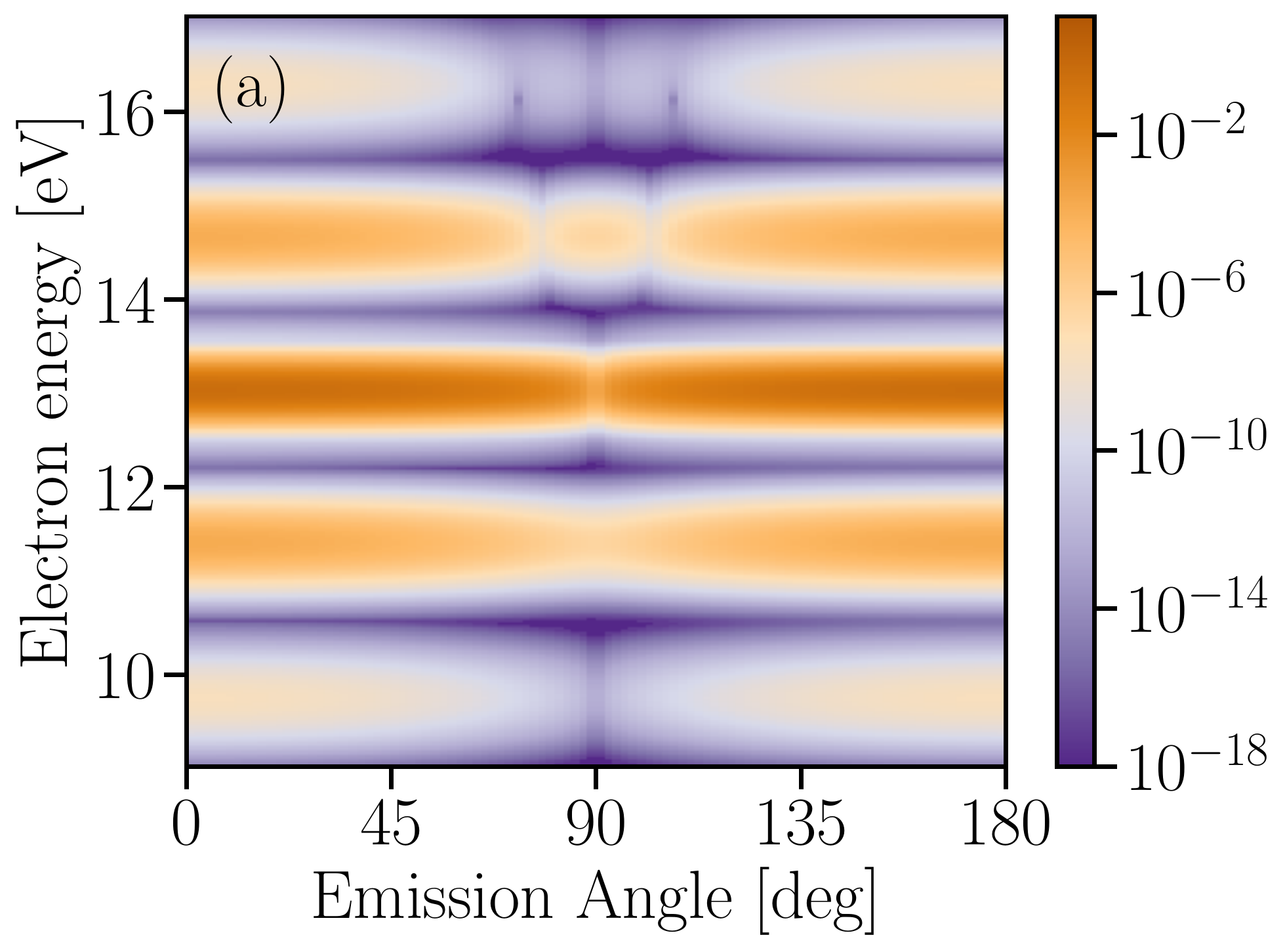}}
  {\includegraphics[scale=0.29]{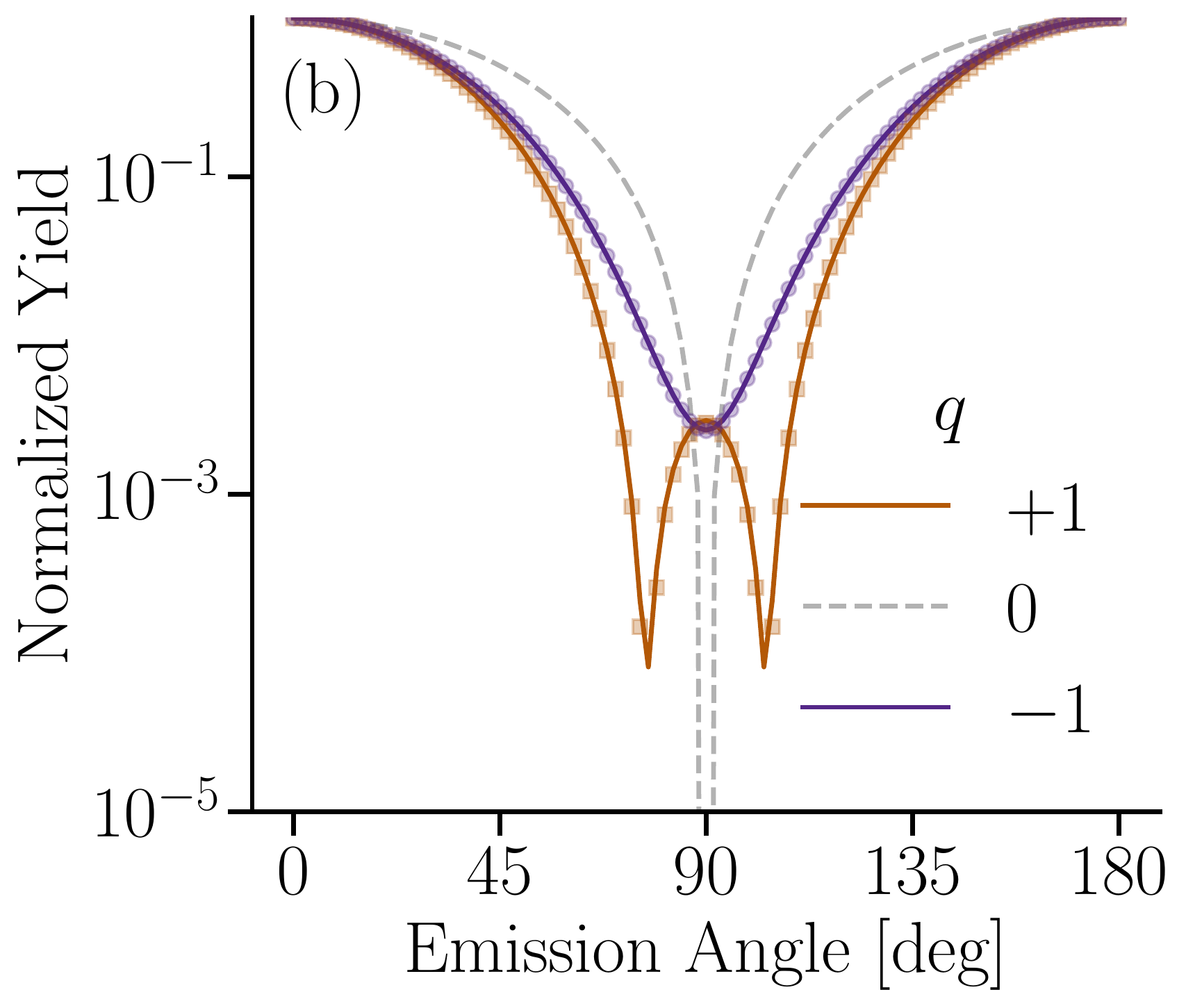}}
  {\includegraphics[scale=0.29]{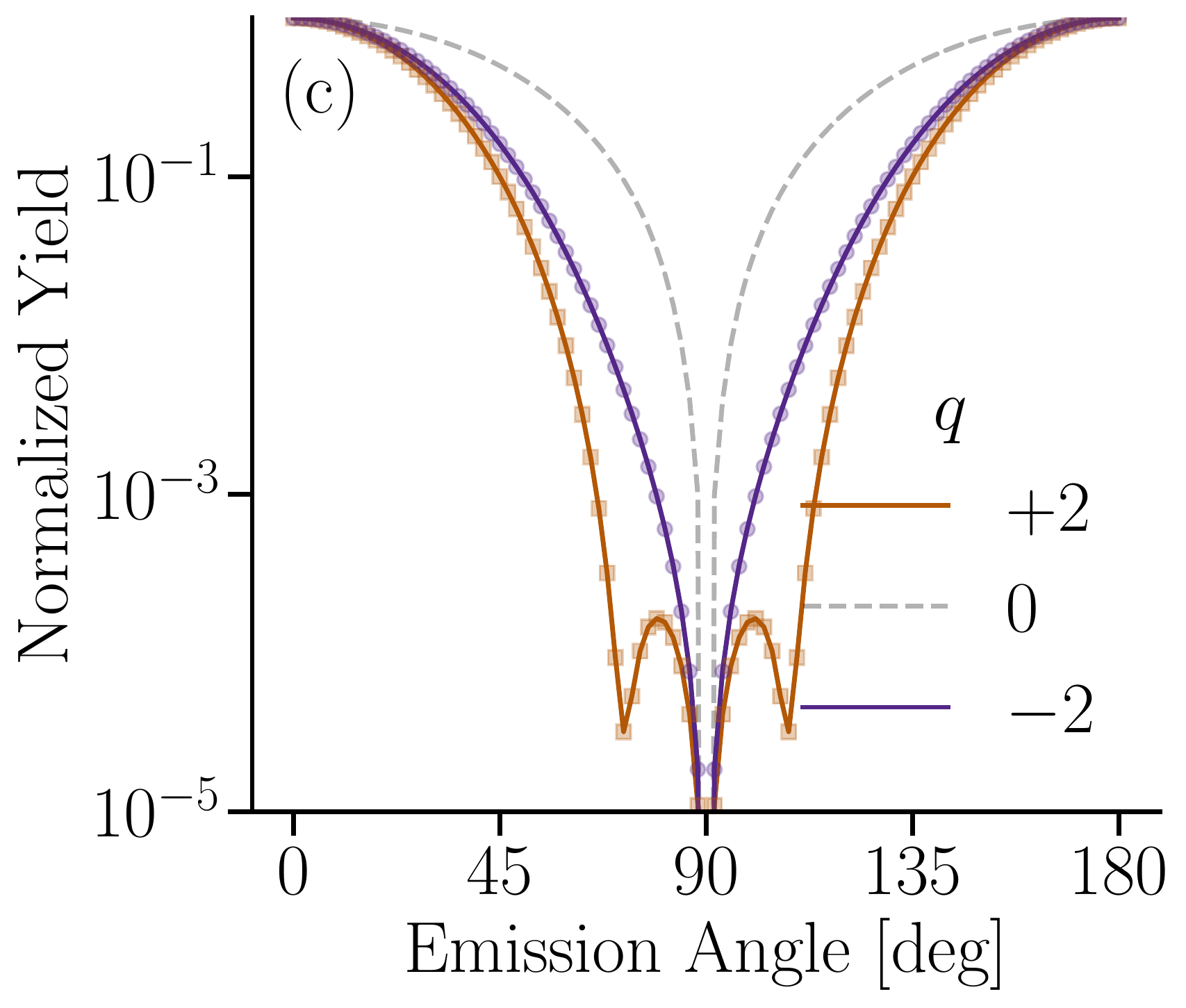}}  
  \caption{(a) Angle-resolved photoelectron spectrum in helium using an XUV-photon energy of 38\,eV. (b) PAD using $q=\pm 1$ and (c) PAD using $q=\pm 2$. The dots in (b,c) are fits to the data using Eq.~\eqref{eq:minimization}.} 
  \label{fig:raw-XUV38-He1s}
\end{figure*}

In Fig.~\ref{fig:raw-XUV38-He1s}~(a) we present our simulation of the angle-resolved photoelectron spectrum in helium on a logarithmic scale. 
The main central line corresponds to absorption of one XUV-photon and the sidebands correspond to absorption or emission of $q$ IR-photons.
Alongside the spectrum, the PADs for sidebands, retrieved by Eq.~\eqref{eq:peak-integration}, are shown in Fig.~\ref{fig:raw-XUV38-He1s}(b) for the first absorption and emission peaks: $q=\pm 1$, and in Fig.~\ref{fig:raw-XUV38-He1s}(c) for the second absorption and emission peaks: $q=\pm 2$.
The main peak, $q=0$, is included for reference. The maxima of the angular distributions are normalized to unity in order to make the comparison between the PADs of the different peaks easier.
The partial wave fitting, where only dipole-allowed spherical harmonics are included in the sum of Eq.~\eqref{eq:minimization}, is in excellent agreement with the simulated PAD for all peaks, $q$.   
The PAD shows an asymmetry between absorption and emission of IR-photons in the continuum which is expressed by the different number of minima in the sideband peaks.
For example, in the first absorption and emission peaks, $q=1$ and $q=-1$, we observe two minima and one minimum, respectively.

\begin{figure}
  \centering
  {\includegraphics[scale=0.45]{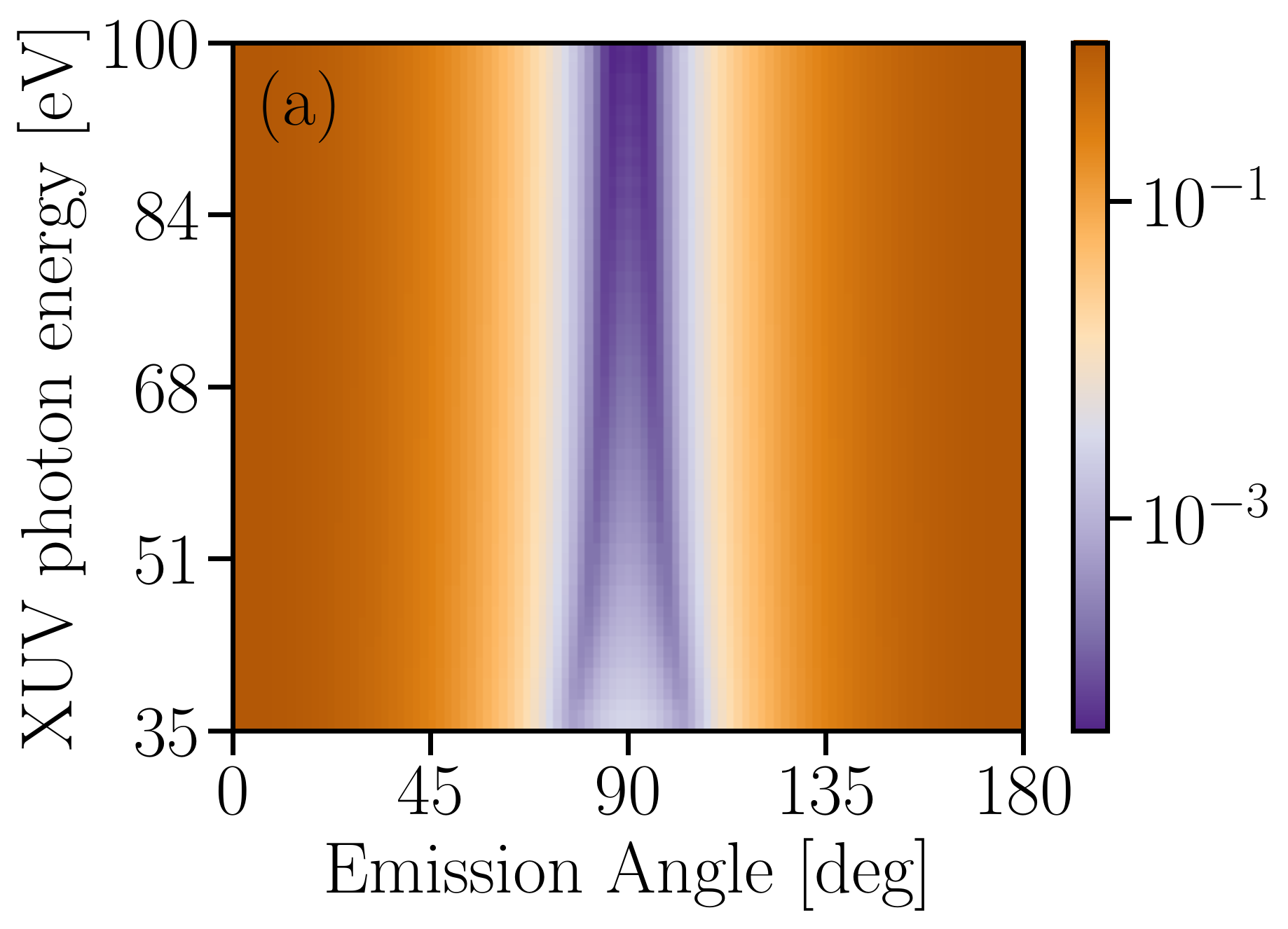}}
  {\includegraphics[scale=0.45]{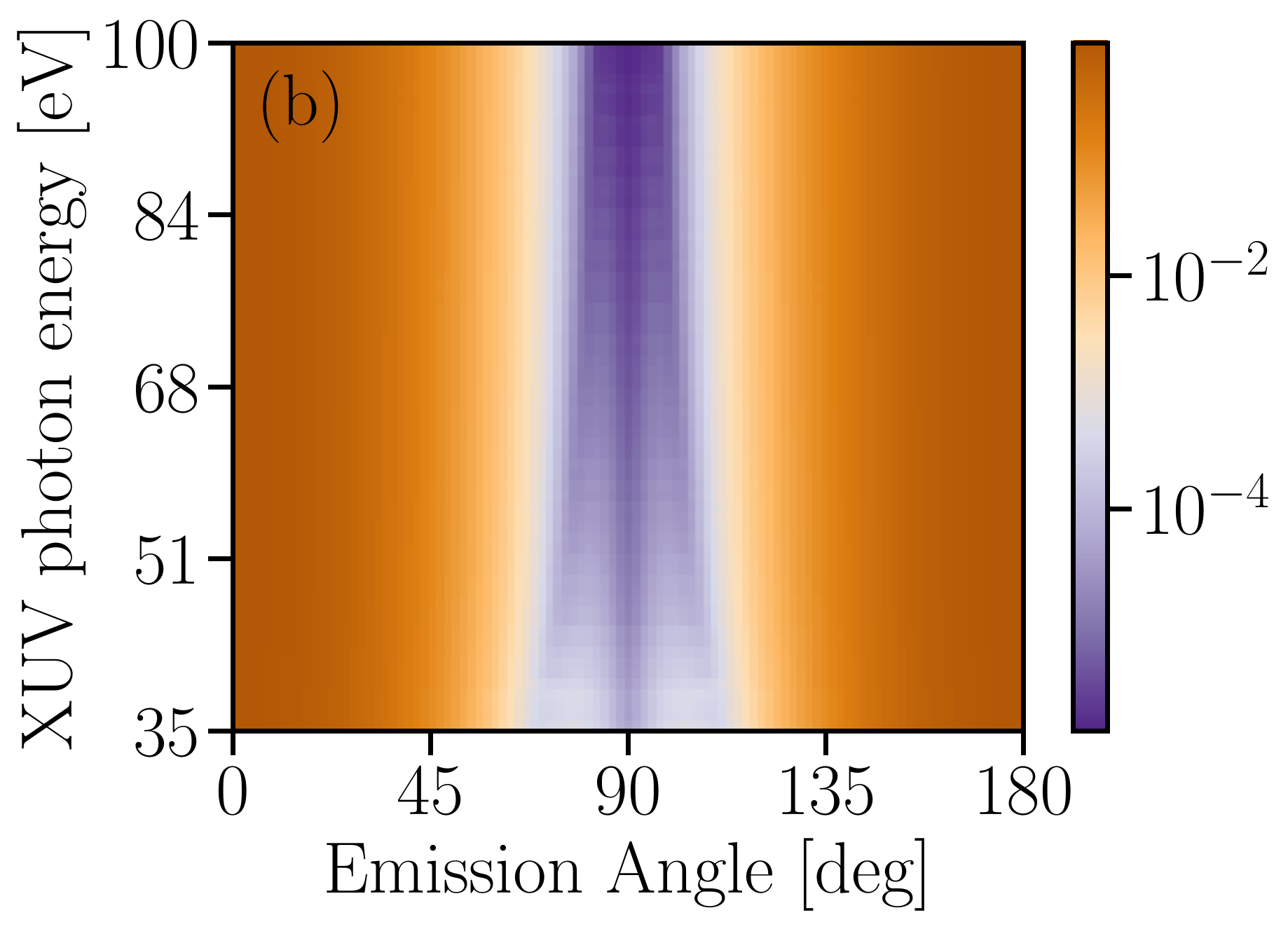}}
  \caption{PAD of helium peaks (a) $q=1$ and (b) $q=2$ as a function of XUV-photon energy.}
  \label{fig:gaffel-He1s}
\end{figure}

In Fig.~\ref{fig:gaffel-He1s}, we present the normalized PAD as a function of XUV-photon energy for $q=1$ and $q=2$ peaks in helium. A uniform filter is applied to smoothen spurious oscillations (that are inherent to the t-SURFF method for Coulomb-like problems). Each figure corresponds to multiple laser-assisted photoionization simulations with all parameters fixed except the frequency of the ionizing XUV-field. In the high kinetic energy limit, the multiple minima of the absorption peaks, $q\in[1,2]$, tend towards a polar  angle of $90$ degrees. 
On the contrary, in the case of emission, the position of the single minimum is independent of the kinetic energy and located at $90$ degrees (not shown). 

\subsection{Neon}
\label{subsec:results-neon}
\begin{figure*}
  \centering
  {\includegraphics[scale=0.29]{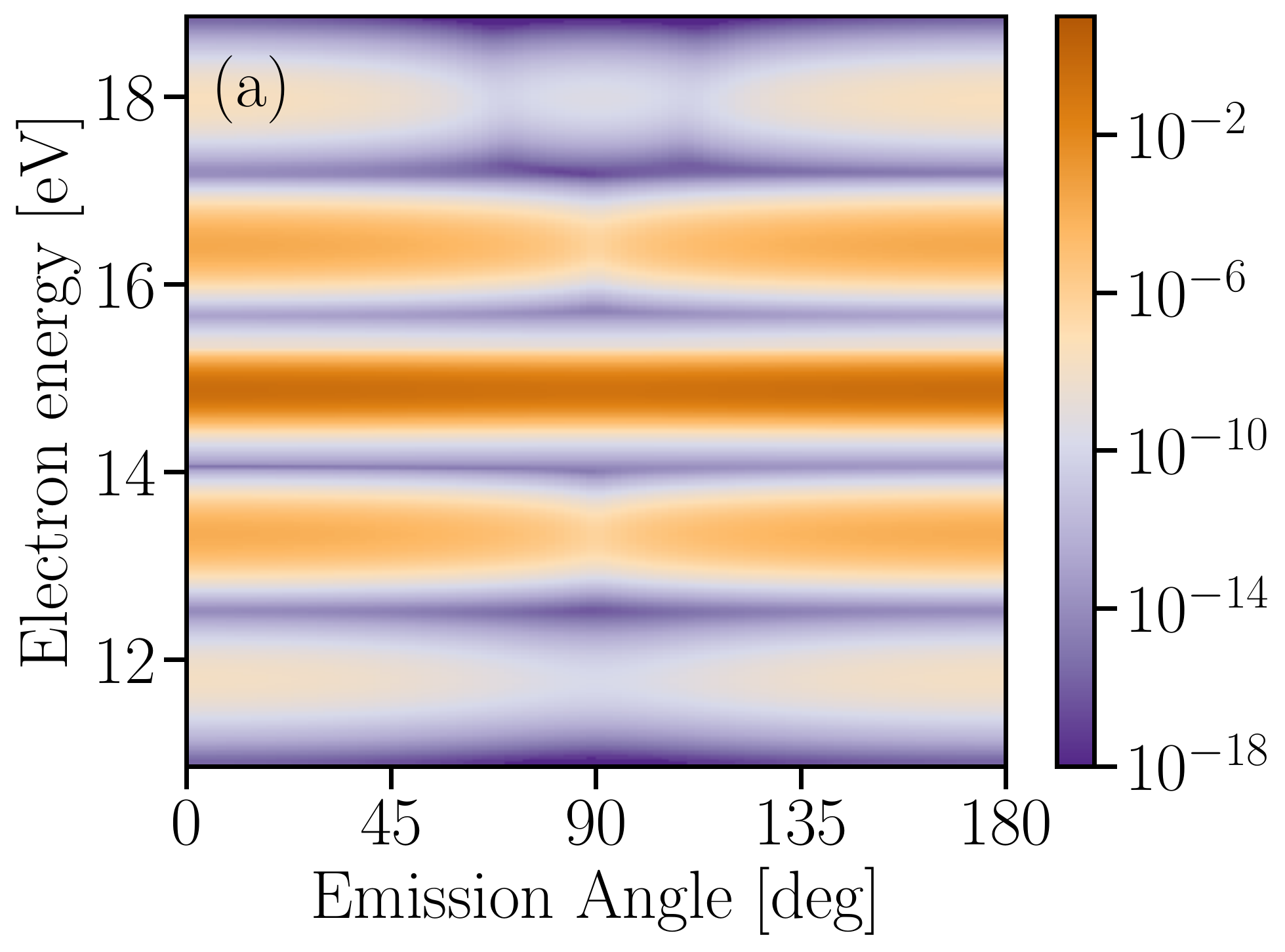}}
  {\includegraphics[scale=0.29]{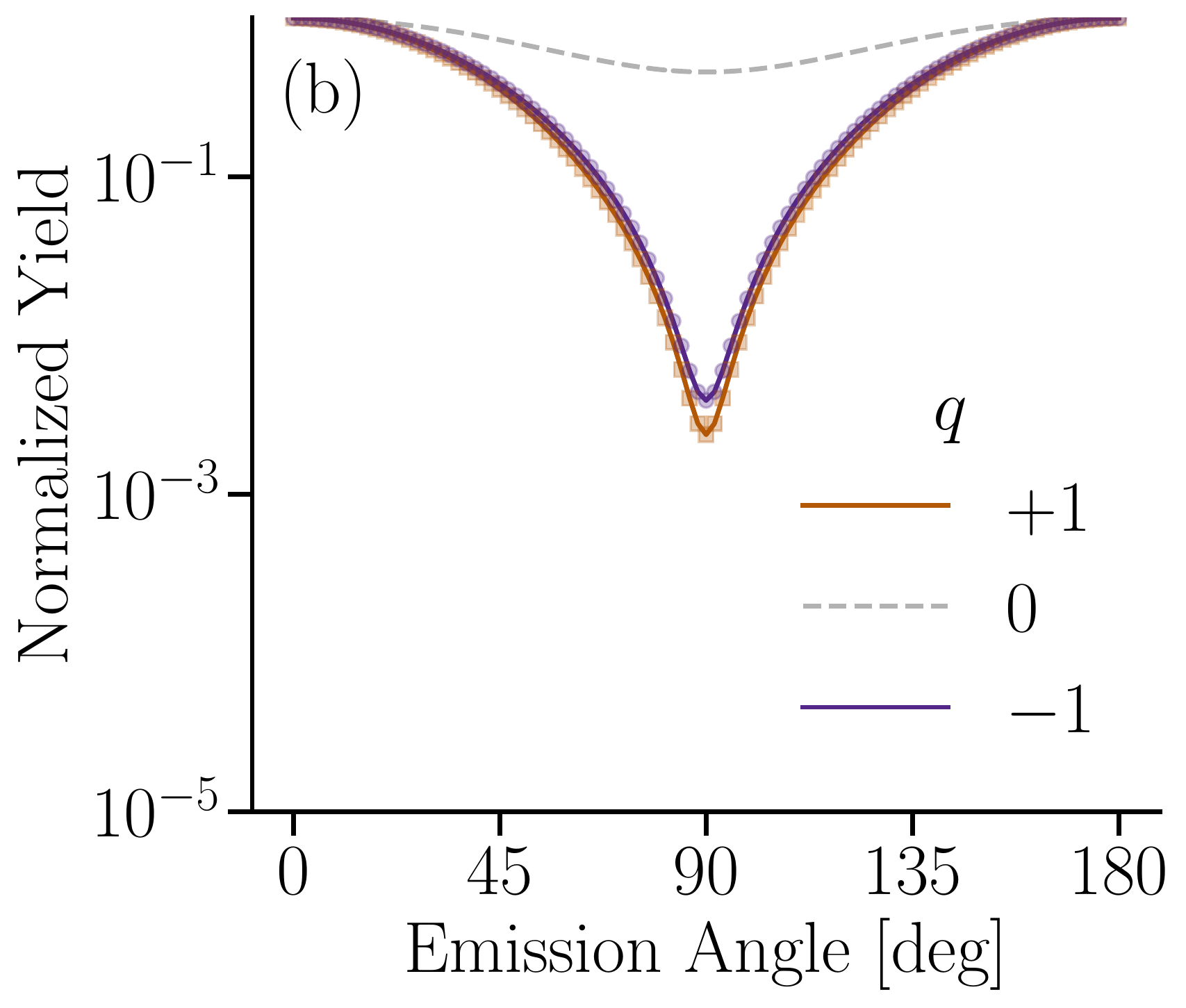}}
  {\includegraphics[scale=0.29]{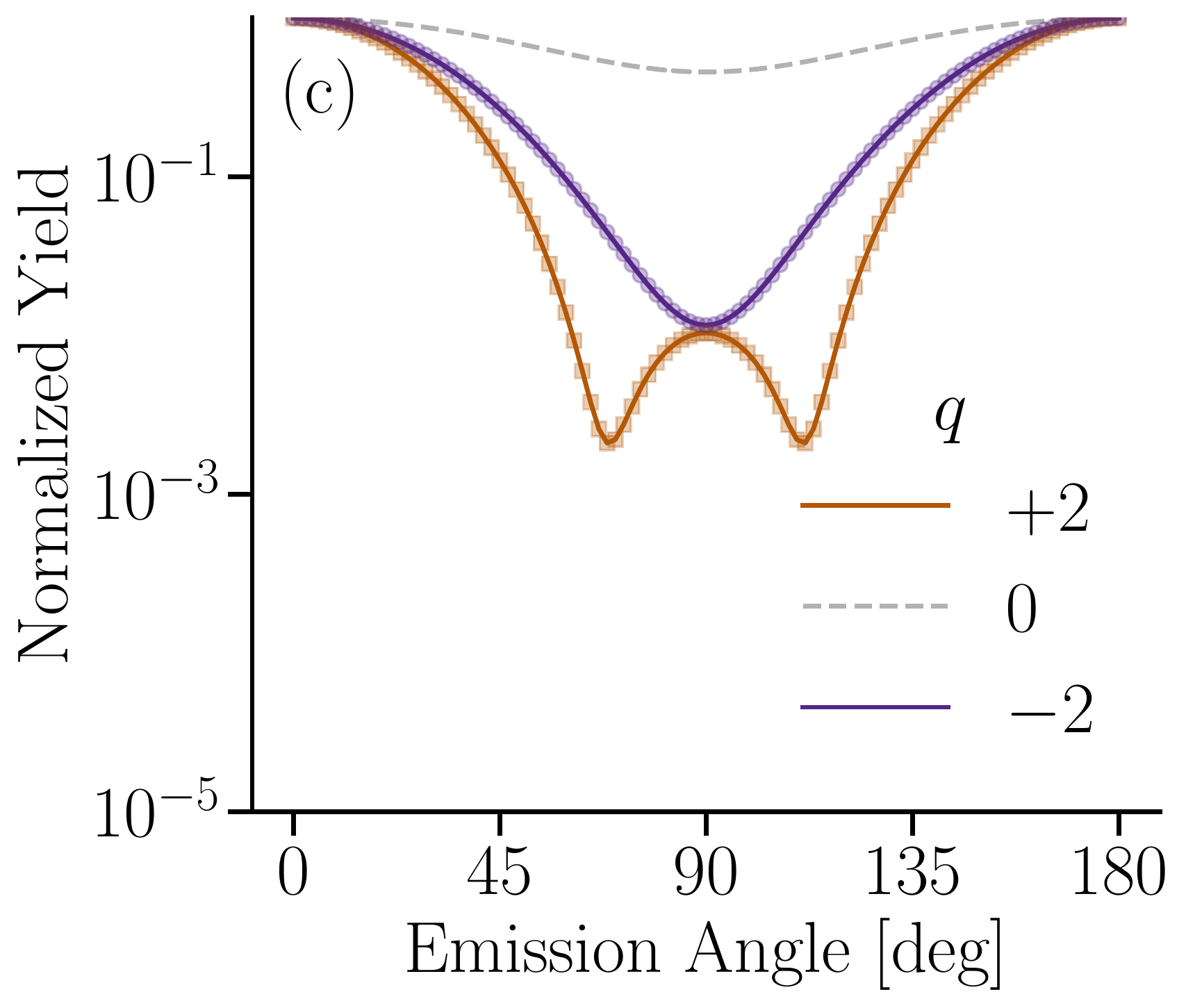}}
  \caption{(a) Angle-resolved photoelectron spectrum in neon $2p$ using an XUV-photon energy of 38\,eV. (b) PAD using $q=\pm 1$ and (c) PAD using $q=\pm 2$. The dots in (b,c) are fits to the data using Eq.~\eqref{eq:minimization}.}
  \label{fig:raw-XUV38-Ne2p}
\end{figure*}

In Fig.~\ref{fig:raw-XUV38-Ne2p}, the obtained angle-resolved photoelectron spectrum in neon (a) is shown alongside the normalized PADs of the first (b) and second (c) absorption and emission peaks.
The fitted spherical harmonics match well with the angular distribution of the peaks of the sidebands.
Contrary to helium, we now have two possible intermediate cases after absorption of an XUV-photon with $\ell=0$ and $\ell=2$.
In neon we do not observe any qualitative difference in the PAD, comparing the first absorption and emission peaks, $q=\pm 1$.
Both peaks show one single minimum with a qualitatively similar angular distribution. 
However, in the second absorption and emission peaks, $q=\pm 2$, there is a clear difference between $q=2$, which shows two distinct minima, and $q=-2$, which shows a single minimum. 

\begin{figure*}
  \centering
  {\includegraphics[scale=0.29]{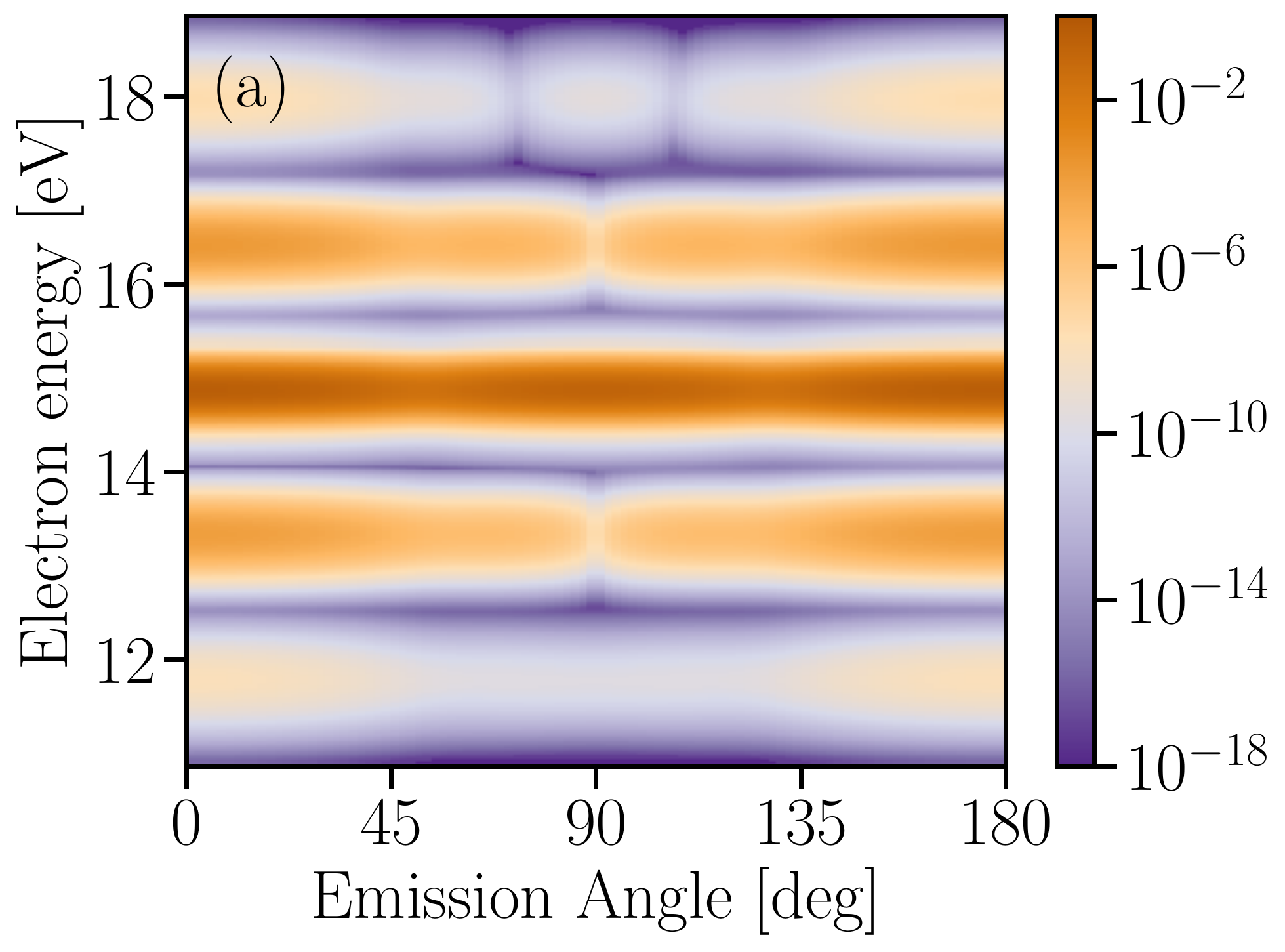}}
  {\includegraphics[scale=0.29]{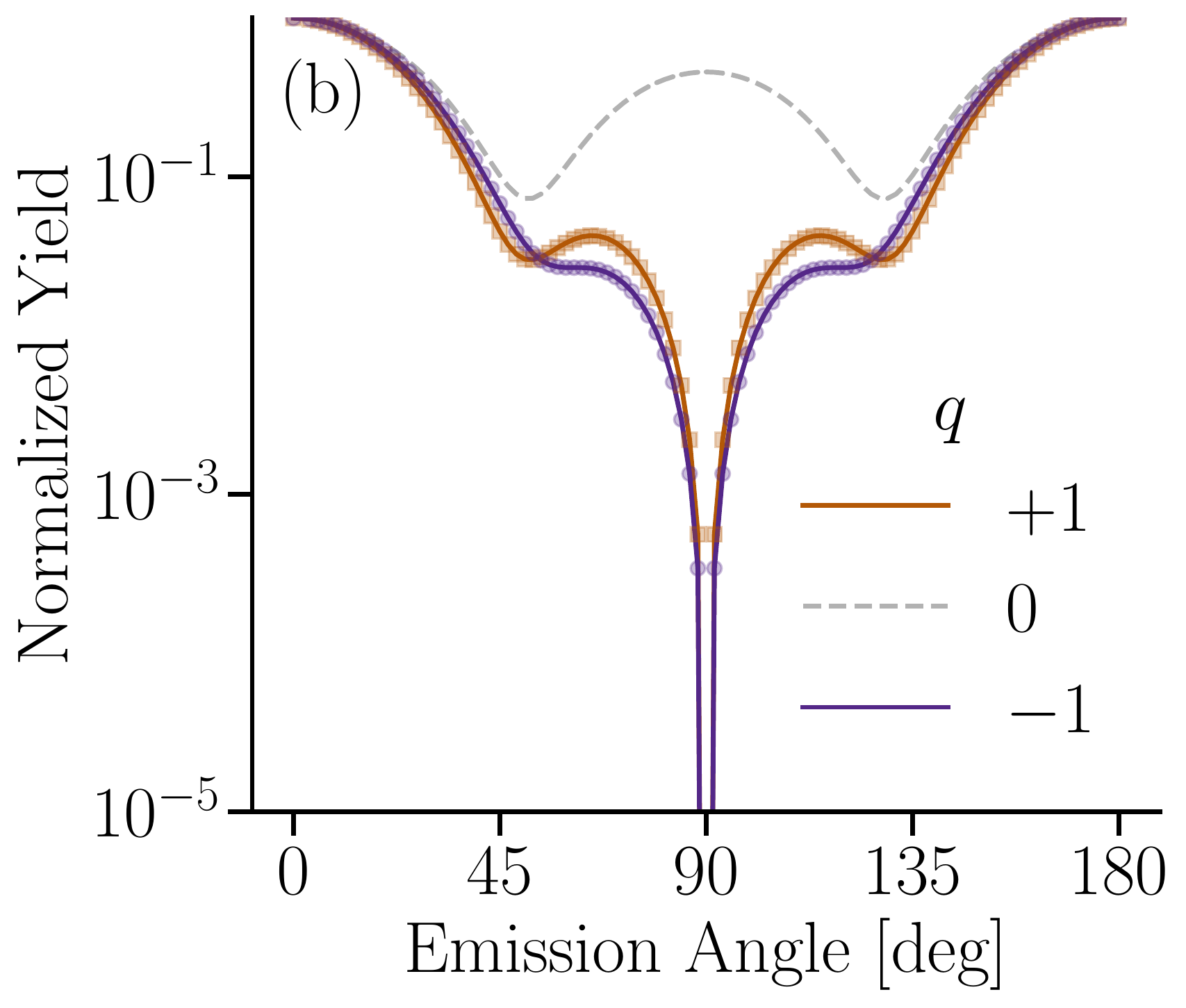}}
  {\includegraphics[scale=0.29]{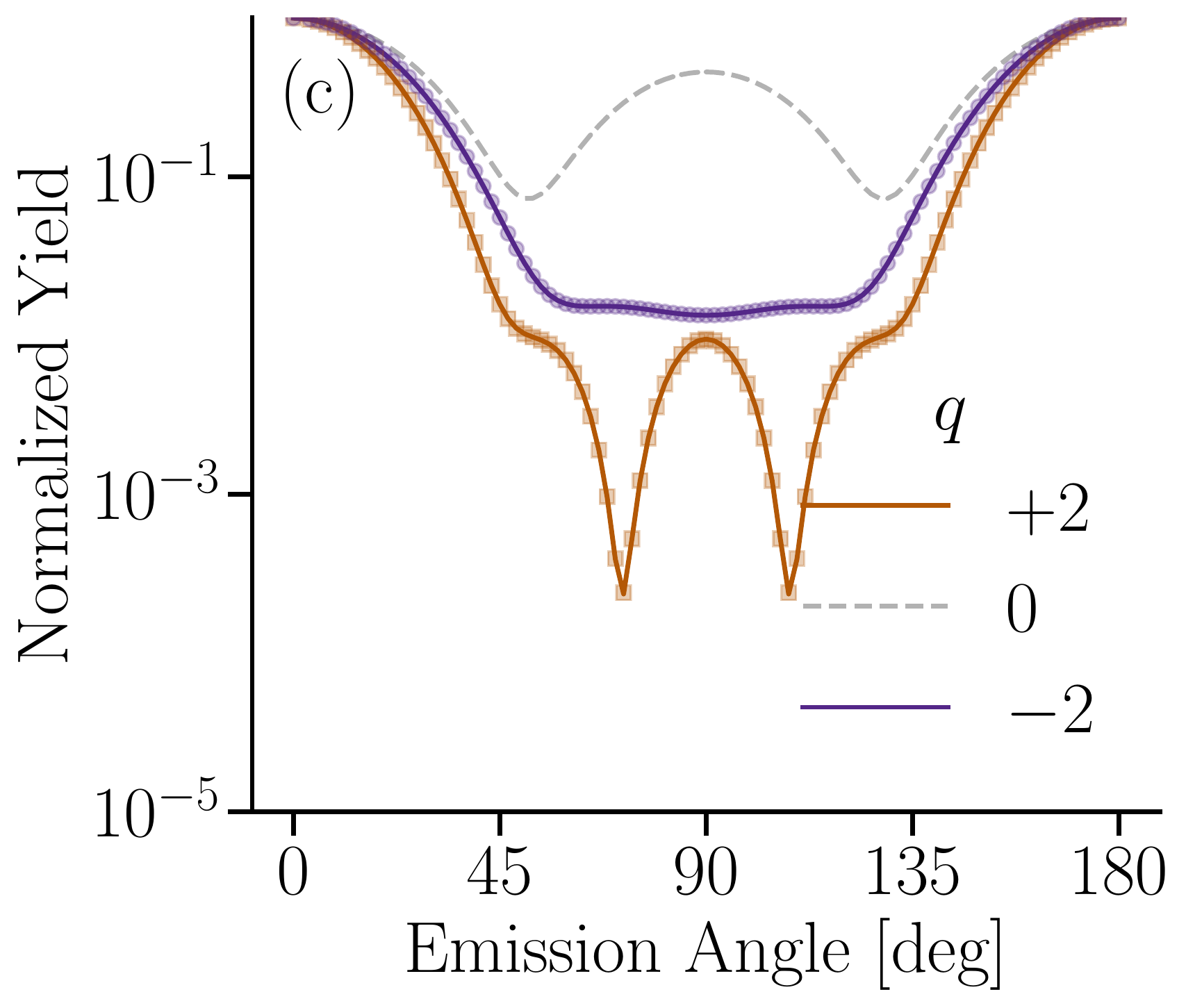}}
  {\includegraphics[scale=0.29]{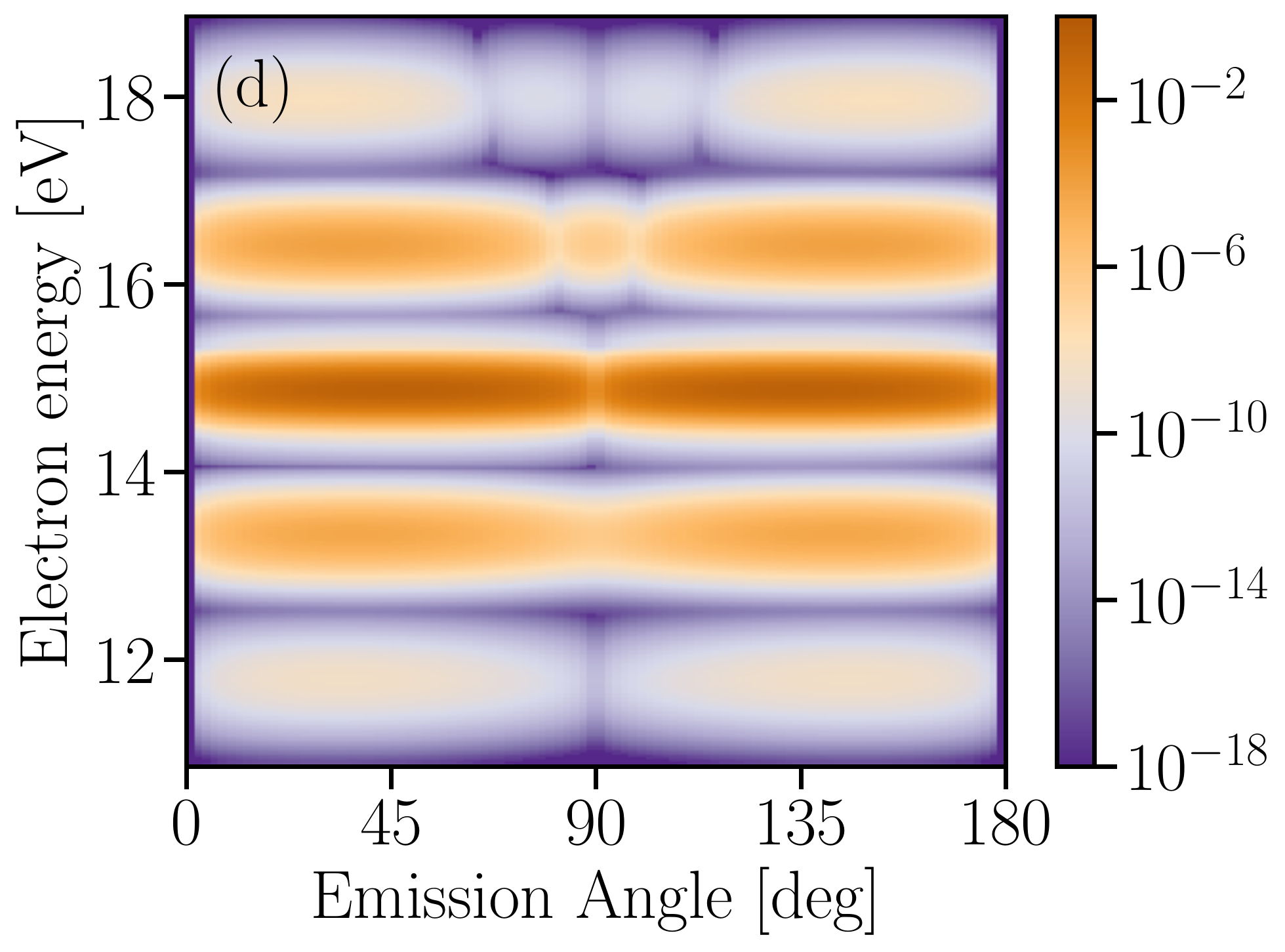}}
  {\includegraphics[scale=0.29]{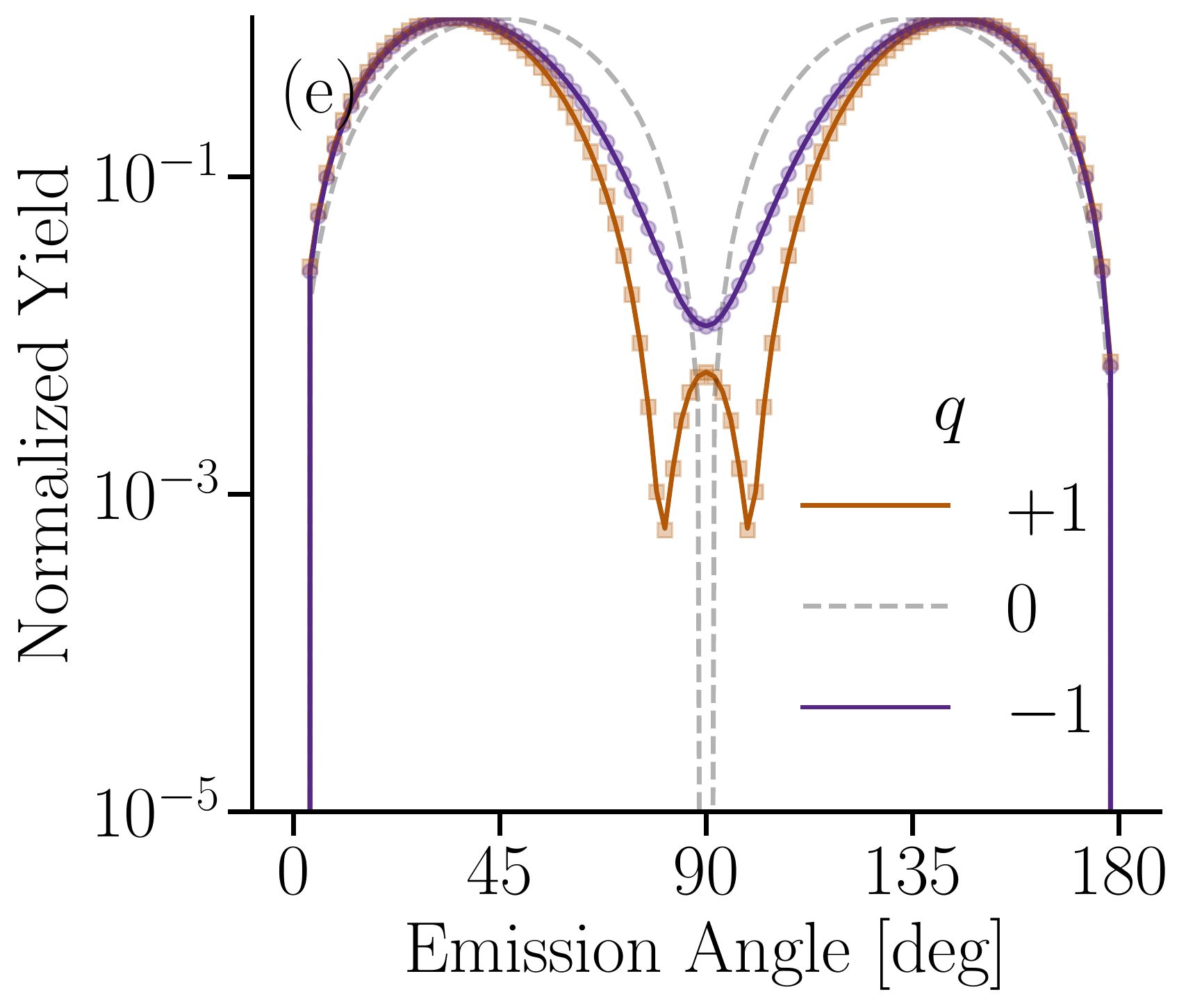}}
  {\includegraphics[scale=0.29]{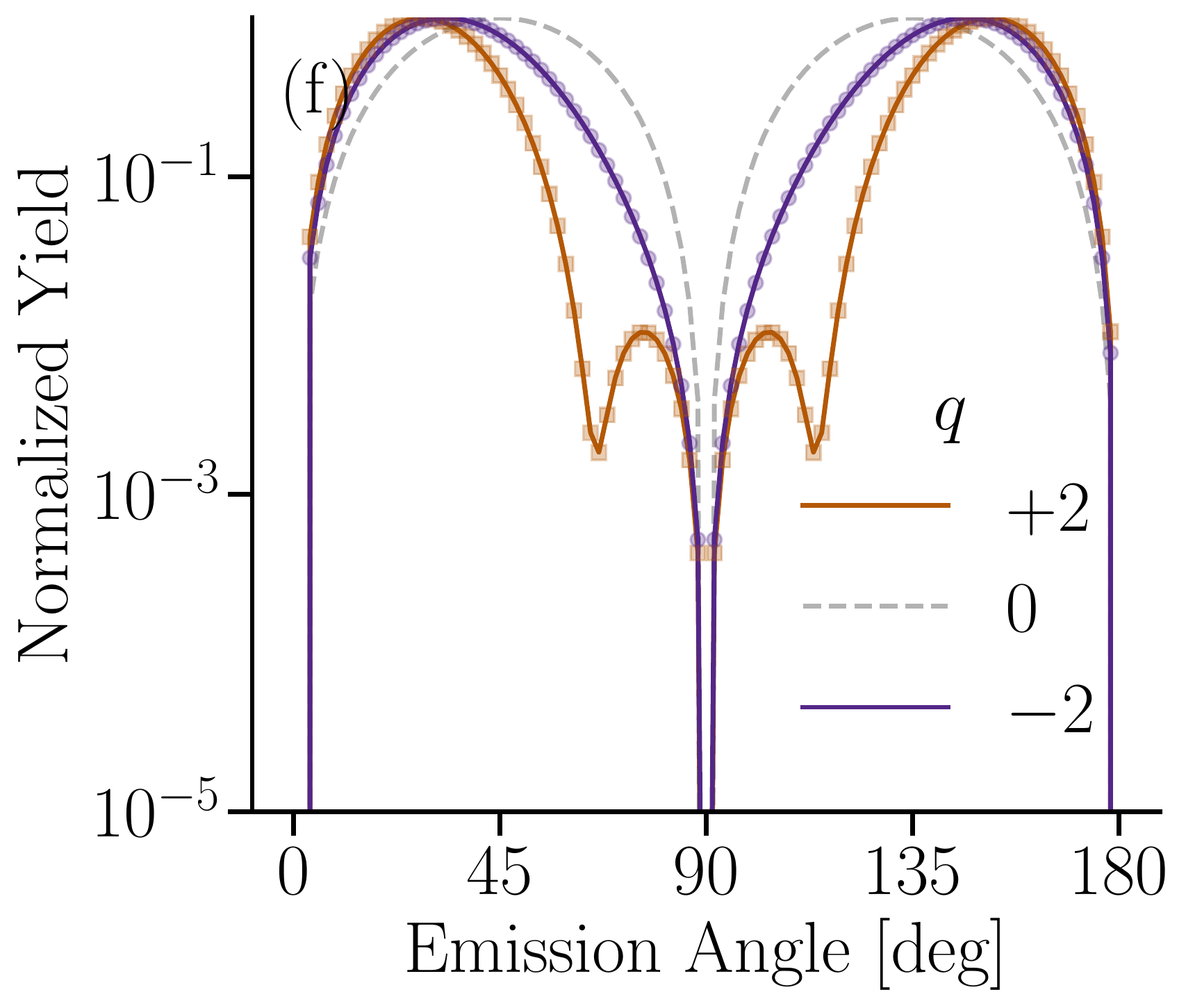}}
  \caption{Angle-resolved photoelectron spectrum and PADs in neon $2p$ with $m=0$ (a-c) and $m=\pm 1$ (d-f). The XUV-photon energy is 38\,eV. The dots in (b,c) and (e,f) are fits to the data using Eq.~\eqref{eq:minimization}.}
  \label{fig:raw-Ne2p-XUV38-m}
\end{figure*}

Unlike helium, the angular distribution in neon results from an incoherent superposition of magnetic quantum numbers of the hole, $m=m_a$. Therefore, we complement our neon studies with $m$-resolved PADs.  
Since we deal with systems of spherical symmetry, the positive $m=+1$ channel and the negative $m=-1$ channel will yield the same photoelectron angular distribution, and without loss of generality we can consider it one effective channel.
We denote this channel as the \textit{gerade} $m=\pm 1$ channel \cite{PabstPRA2012}. 
In Fig.~\ref{fig:raw-Ne2p-XUV38-m}, the angle-resolved photoelectron spectra are presented in neon on a logarithmic scale for $m=0$ and $m=\pm 1$, respectively, alongside the normalized PAD of the absorption and emission peaks.
For the $m=0$ channel isolated, the PADs of $q=\pm 1$ show small  differences. In absorption we find with two shallow minima, while in emission there are instead two shoulders. Both absorption and emission have a deep minimum at 90 degrees. In total the $m=0$ case has three minima for $q=1$, and one minimum for  $q=-1$. 
In the $m=\pm 1$ channel, the difference between one-photon absorption and emission, $q=\pm 1$, is more distinct because the PAD shows four and three clear minima, respectively (including the minima at 0 and 180 degrees). 
In $q=\pm 2$, there is a clear difference between absorption and emission of two IR photons for both $m=0$ and $m=\pm 1$. 
In the $m=0$ channel, we identify two minima and two outer shoulders in the case of absorption, $q=2$, and a flat region with a single shallow minimum in the case of emission, $q=-2$. 
Likewise, in $m=\pm 1$, we identify a clear difference between $q=2$ and $q=-2$. The $q=2$ peak has five minima, while the $q=-2$ peak has three minima (including the minima at the polar angle of 0 and 180 degrees).  

\begin{figure*}
  \centering
  {\includegraphics[scale=0.29]{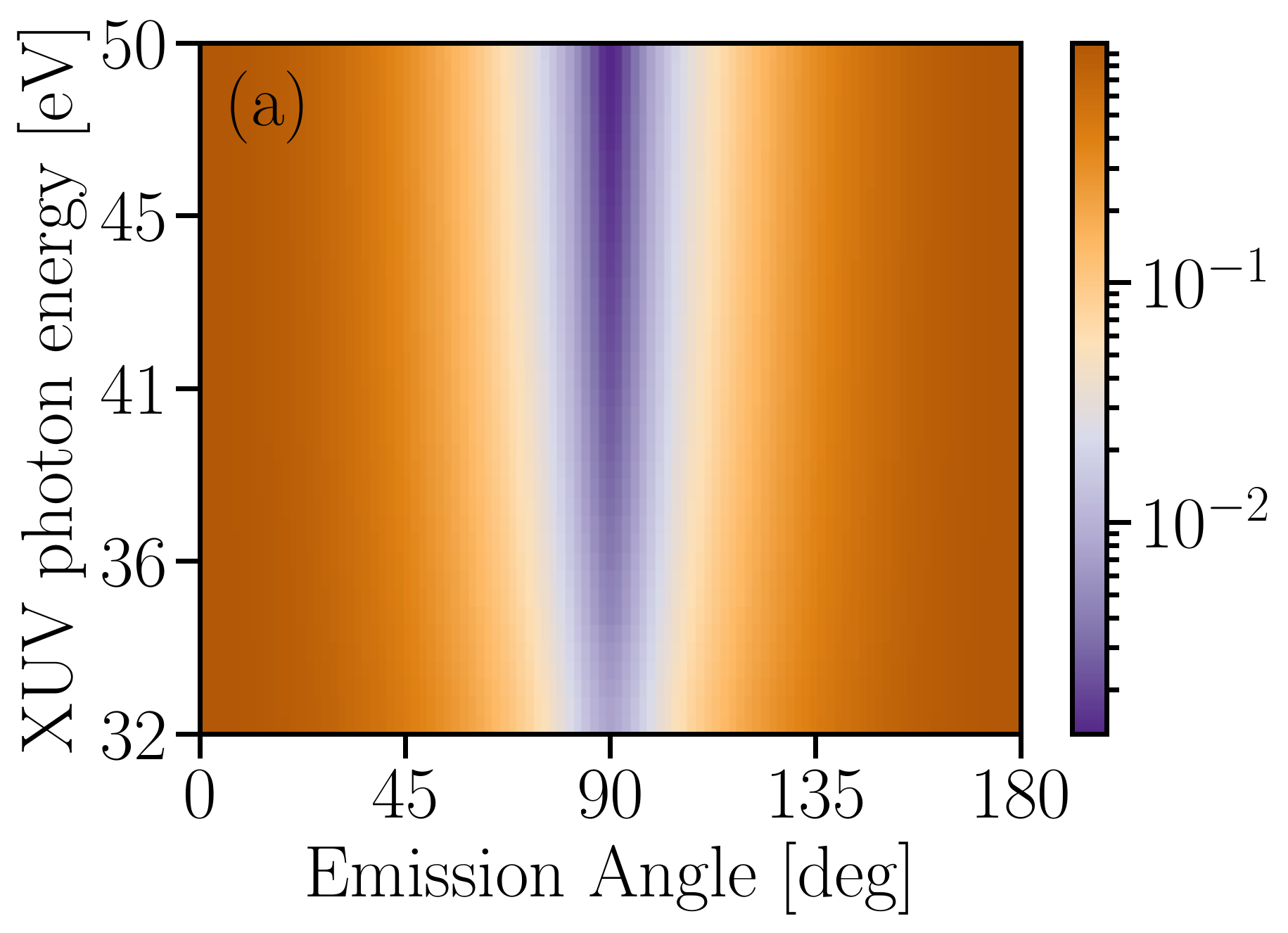}}
  {\includegraphics[scale=0.29]{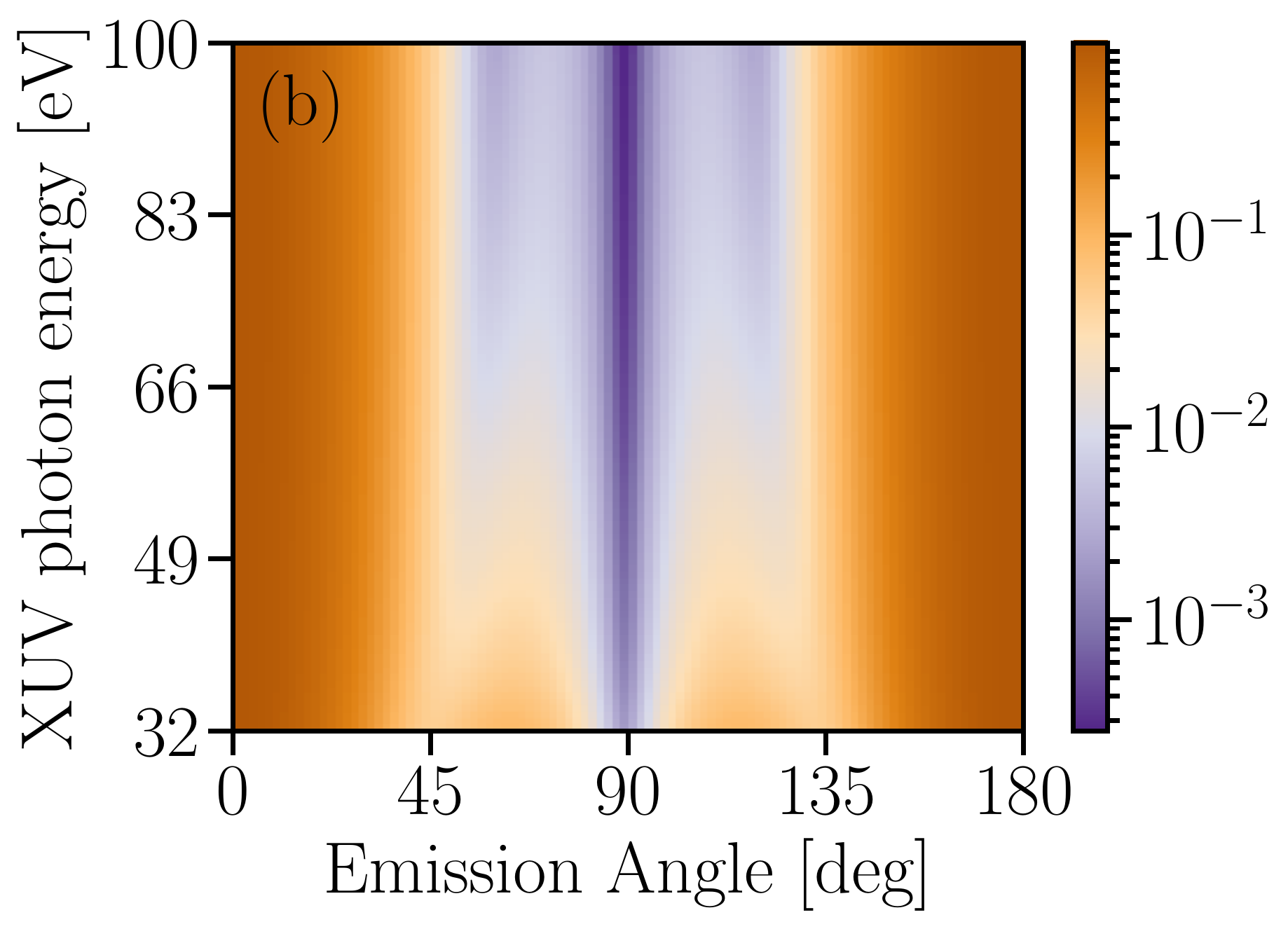}}
  {\includegraphics[scale=0.29]{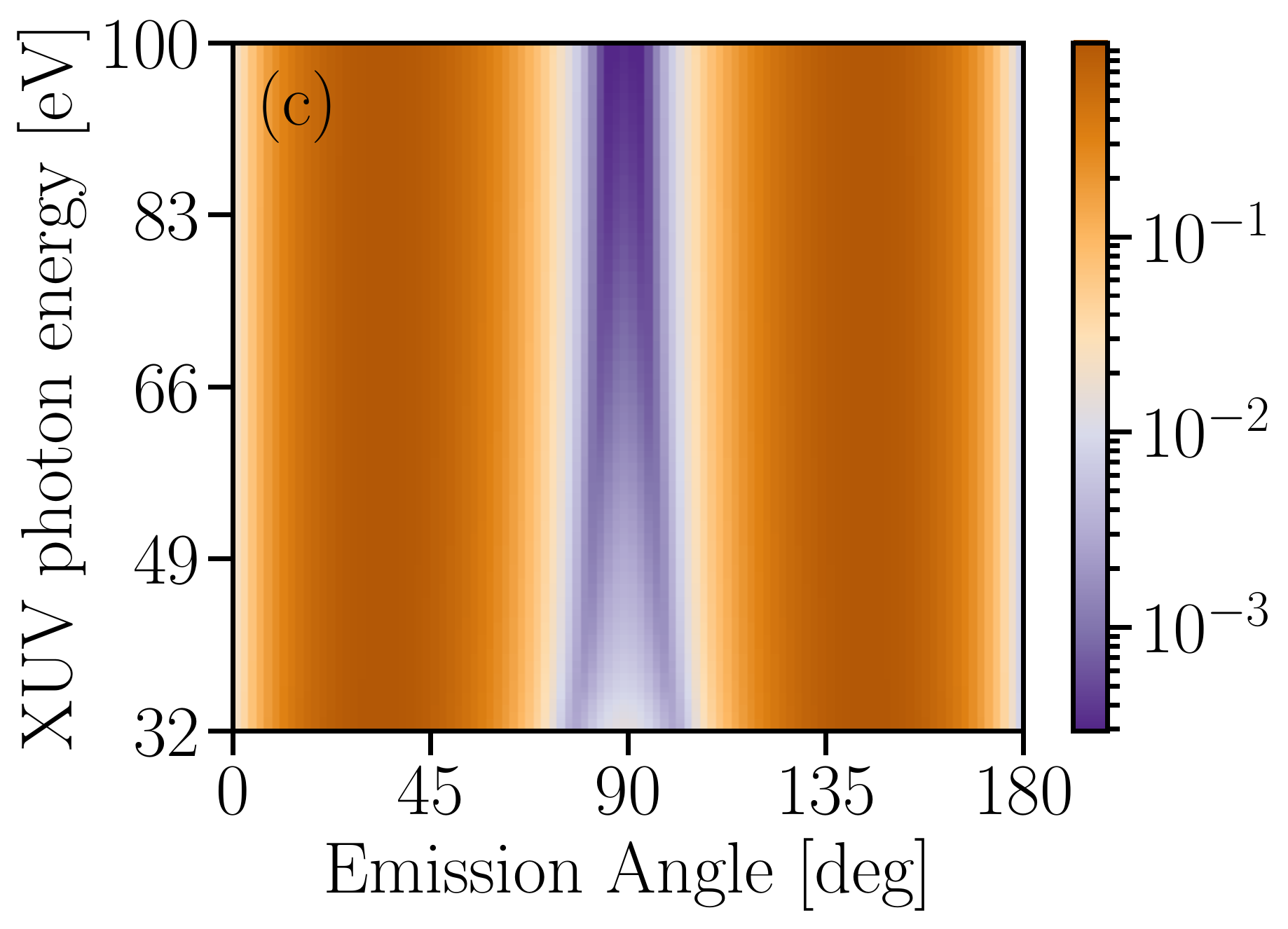}}
  {\includegraphics[scale=0.29]{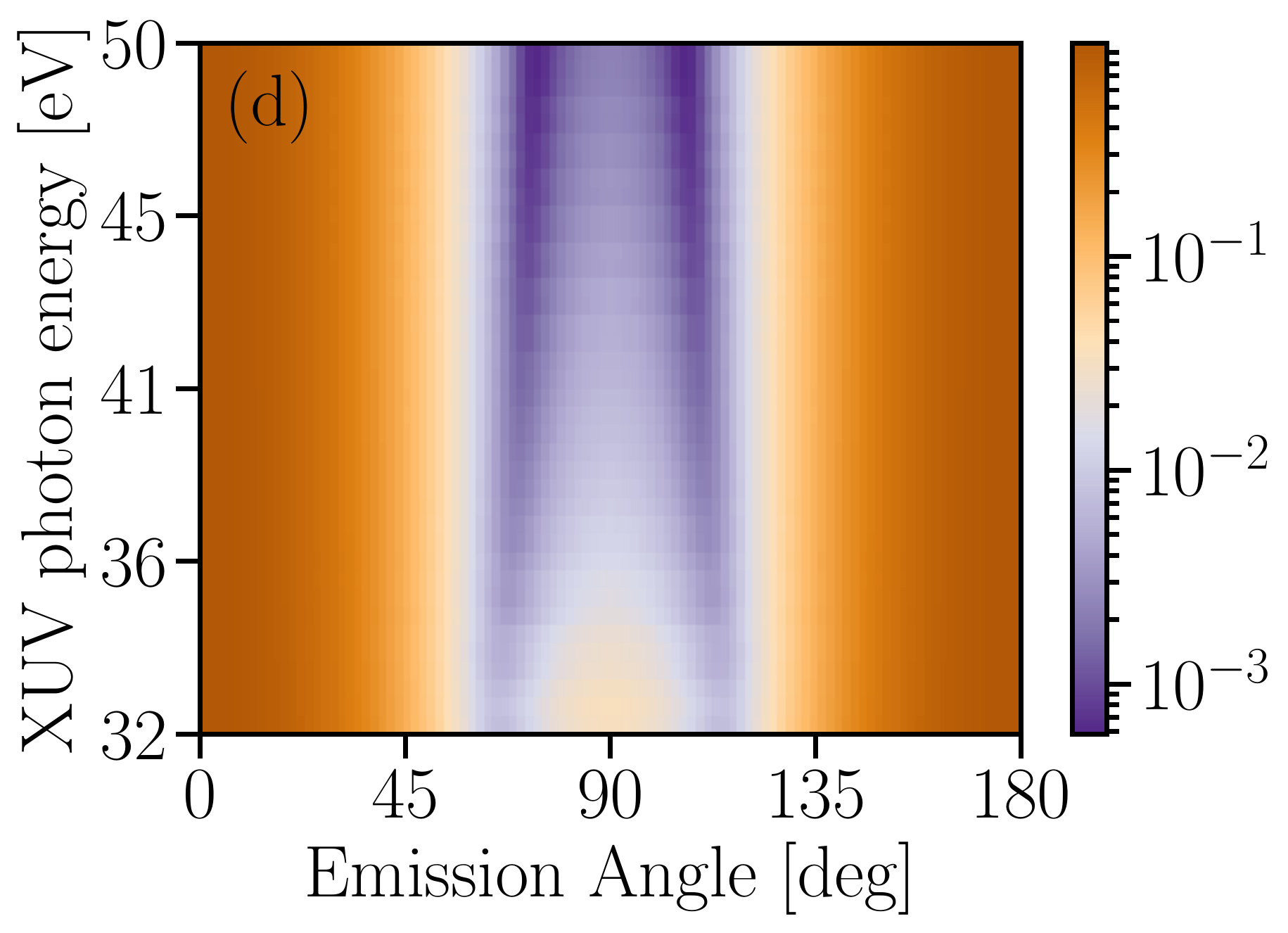}}
  {\includegraphics[scale=0.29]{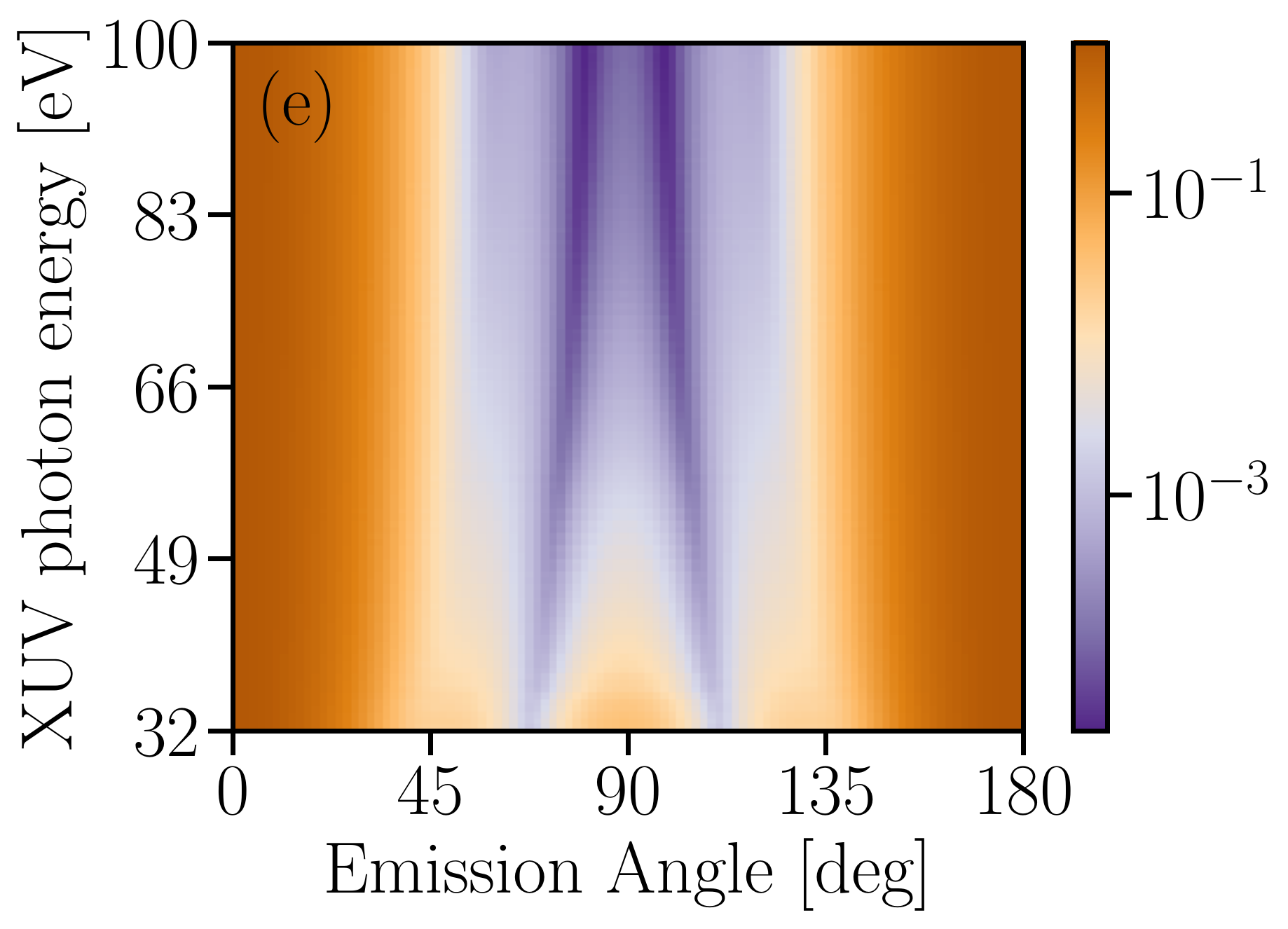}}
  {\includegraphics[scale=0.29]{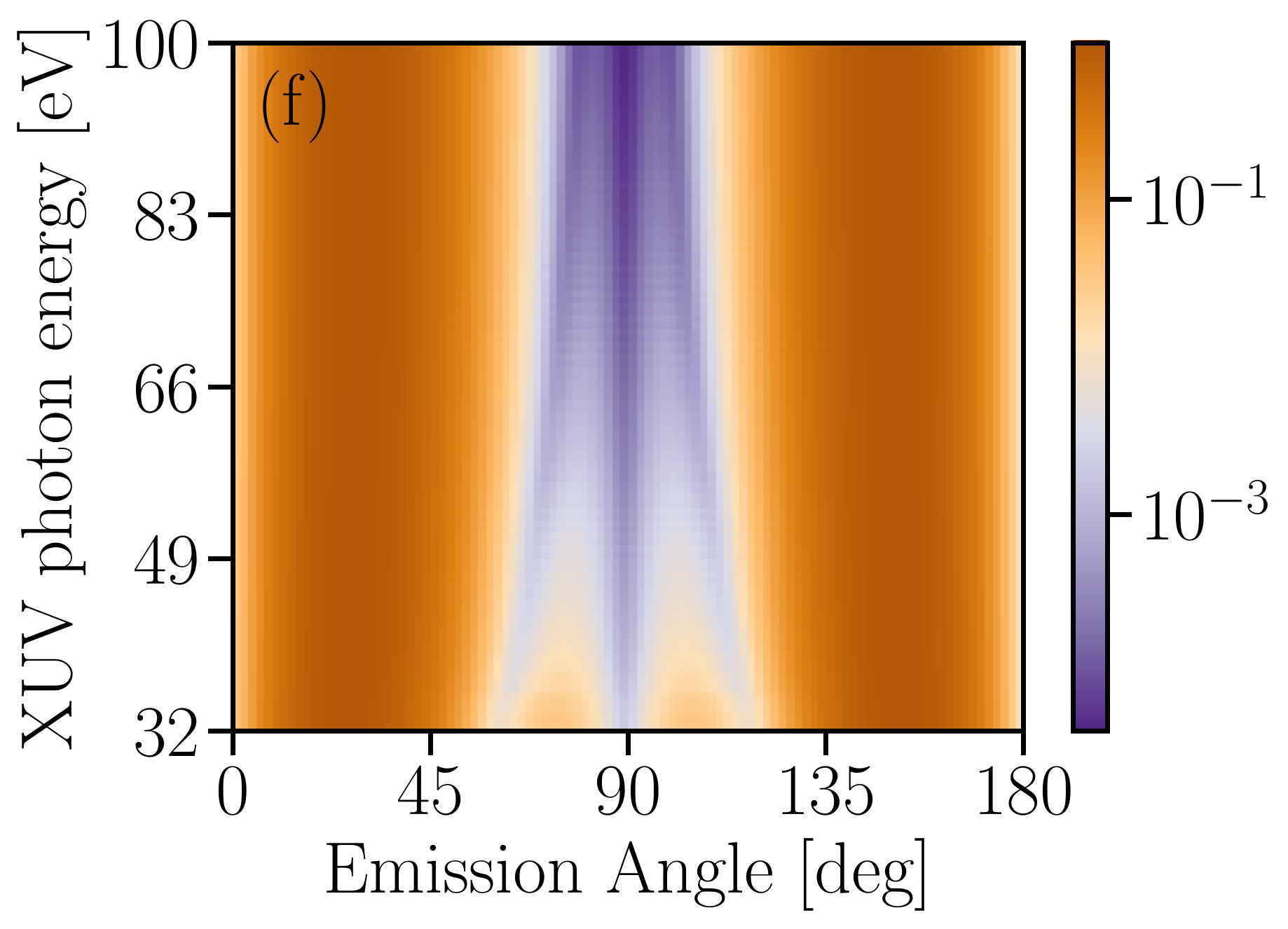}}
  \caption{PAD of neon as a function of XUV-photon energy for peaks (a) $q=1$ unresolved in $m$, (b) $q=1$ with $m=0$, (c) $q=1$ with $m=\pm 1$, (d) $q=2$ unresolved in $m$, (e) $q=2$ with $m=0$ and (f) $q=2$ with $m=\pm 1$.}
  \label{fig:gaffel-Ne2p}
\end{figure*}

In Fig.~\ref{fig:gaffel-Ne2p}, we present the normalized PAD of peaks $q=1$ and $q=2$ as a function of the XUV-photon energy for neon in the non-resolved $m$ case, and the two resolved $m=0$ and $m=\pm 1$  cases. 
While the PADs change in shape with increasing XUV photon energy, they maintain their qualitative attributes. We note that the neon $m=\pm 1$ cases, shown in Fig.~\ref{fig:gaffel-Ne2p}~(c) and (f), are qualitatively similar to the helium $m=0$ case, shown in Fig.~\ref{fig:gaffel-He1s}~(a) and (b), with the exception of two stationary minima in neon at 0 and 180 degrees. 

\section{Discussion}
\label{sec:Discussions}
In the previous section we have shown that the PADs of sidebands due to absorption and emission of IR photons exhibit qualitative differences. In order to understand this difference we turn to a partial wave analysis that allows us to study the relative strength of different laser-assisted photoionization paths (see Fig.~\ref{fig:sketch}). A similar approach was recently used by Busto~et al.\ to study the first-order sidebands, $q=\pm1$~\cite{busto_fano_2019}.      
Each transition between partial waves in the continuum is determined by a radial dipole integral as well as an angular dipole integral. 
According to Fano's propensity rule for photoionization \cite{fano_propensity_1985}, the radial integral favours transitions to higher angular momentum. In the continuum the photoelectron can both absorb and emit IR photons, which respectively favours increasing and decreasing angular momentum \cite{busto_fano_2019}. This is a direct consequence of time-reversal symmetry for the two continuum processes. 
In the high-energy limit, this radial effect vanishes, while the angular effects remain constant. 
This implies that the branching ratio of different partial waves will be determined by the angular integrals for both absorption and emission sidebands in the high-energy limit. 

In order to study unique partial-wave paths in the continuum to the first sidebands, $q=\pm 1$, we consider helium with $m=0$ and neon with $m=\pm 1$. These are special transitions because they only have one intermediate angular momentum that is reached after absorption of the XUV-photon. Therefore, it is easy to compare one-photon IR absorption and emission processes directly using the complex amplitudes of partial waves, $\tilde{a}_{\ell_{>/<}}^{\pm q}$, extracted by Eq.~\eqref{eq:minimization} for $q=\pm 1$. The notation,  $\ell_{>/<}$, refers to increasing and decreasing angular momentum,  respectively, as defined above Eq.~\eqref{eq:fthetavarhpi}. 

\begin{figure}
  \centering
  \includegraphics[scale=0.35]{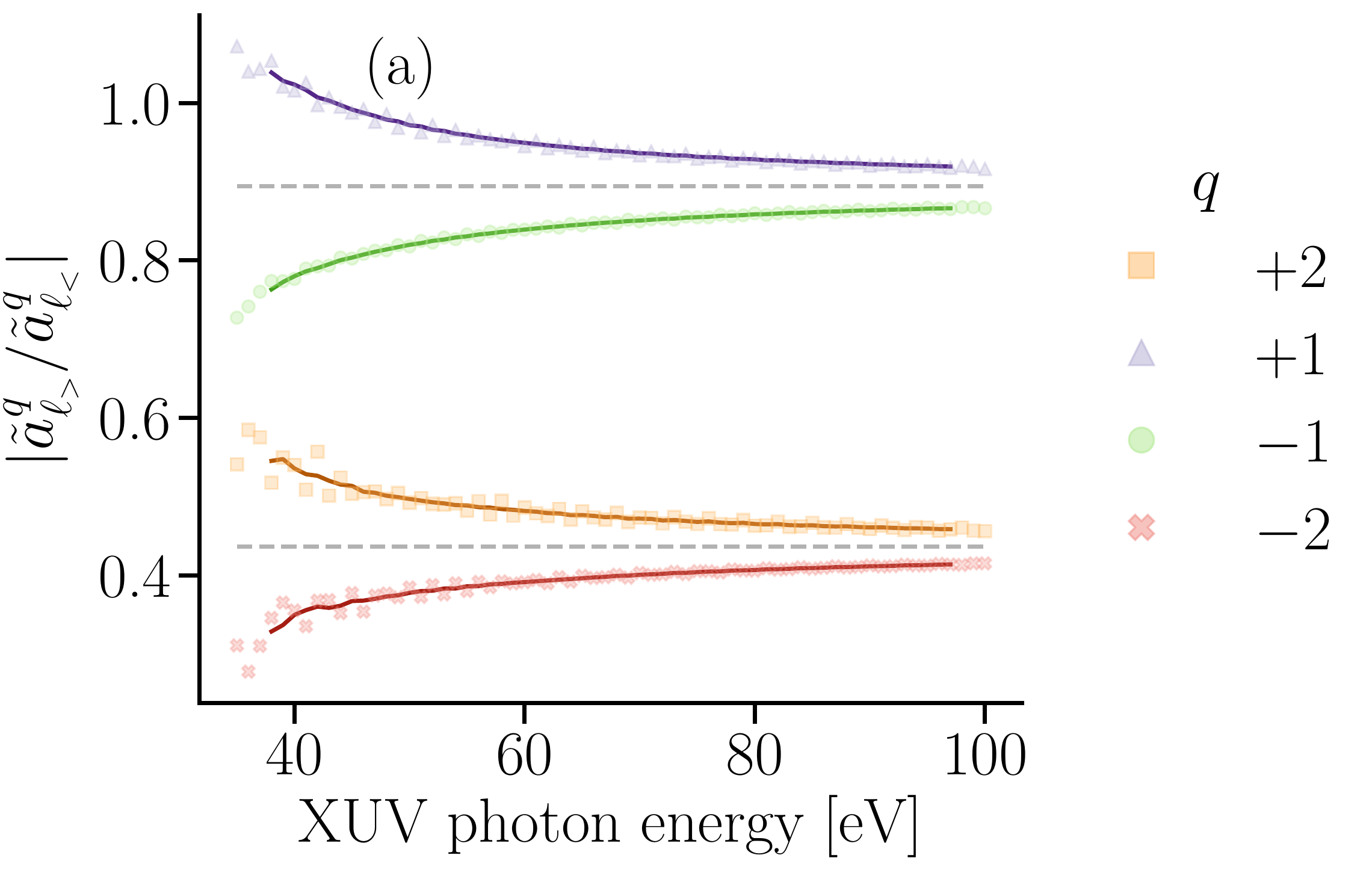}
  \includegraphics[scale=0.35]{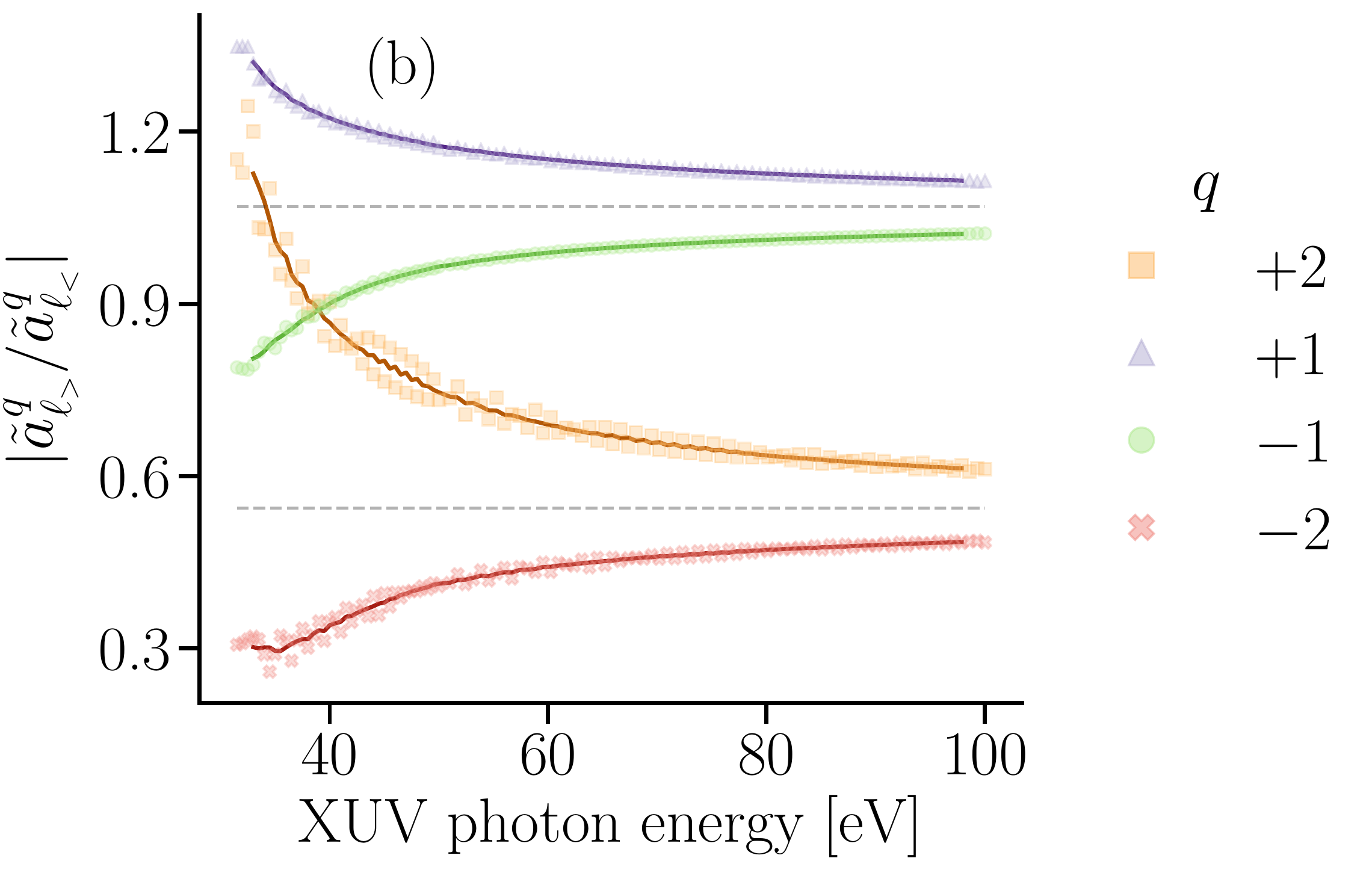}
  \caption{Ratio of the magnitude of the coefficients of fitted spherical harmonics of the peaks of the angle-resolved photoelectron angular distribution as a function of XUV-photon energy in (a) helium and (b) neon, $m=\pm 1$.}
  \label{fig:crest-peaks-compare}
\end{figure}

In Fig.~\ref{fig:crest-peaks-compare} we present the absolute ratio of the complex amplitudes, $|\tilde{a}_{\ell_>}^{\pm q}/\tilde{a}_{\ell_<}^{\pm q}|$. 
In the high-energy limit, we find that the ratios for $q=\pm 1$  approach a value determined by the angular part of the dipole matrix element, shown in Fig.~\ref{fig:crest-peaks-compare} as a gray dotted line.
In helium, for $q\pm1$, the limit of the ratio is given by
\begin{equation}
    \label{eq:one-photon-angular-ratio-he}
    \bigg{|}\frac{\tilde{a}_{\ell_>}^{\pm 1}}{\tilde{a}_{\ell_<}^{\pm 1}} \bigg{|} \to \bigg{|} \frac{\mel{Y_{20}}{Y_{10}}{Y_{10}}}{\mel{Y_{00}}{Y_{10}}{Y_{10}}}\bigg{|} = \frac{2}{\sqrt{5}}, 
\end{equation}
which means that it is more probable to decrease angular momentum. 
At low kinetic energies we find that the absorption process, $q=1$, favours increasing angular momentum due to an enhancement from the radial dipole contribution.   
In neon, for $m=\pm1$ and $q=\pm 1$, the limit is  
\begin{equation}
    \label{eq:one-photon-angular-ratio-ne}
    \bigg{|}\frac{\tilde{a}_{\ell_>}^{\pm 1}}{\tilde{a}_{\ell_<}^{\pm 1}}\bigg{|} \to \bigg{|} \frac{\mel{Y_{31}}{Y_{10}}{Y_{21}}}{\mel{Y_{11}}{Y_{10}}{Y_{21}}} \bigg{|} = \sqrt{\frac{8}{7}}, 
\end{equation}
which means that it is more probable to increase angular momentum. 
At low kinetic energies we find that the emission process, $q=-1$, favours decreasing angular momentum due enhancement from the radial dipole contribution. 
Qualitatively, we understand that the $q=\pm1$ ratios are close to one because there is one unique path to reach each final partial wave. 
This is related to the comparable magnitude of dipole matrix elements from $\ell_0$ to $\ell_0 \pm 1$. 

In the case of $q=\pm 2$ the physics is more complicated because there are two interfering paths leading to the lower angular momentum, while there is one unique path to the higher angular momentum for helium with $m=0$ and neon with $m=\pm 1$. The two interfering paths leading to the lower angular momentum is coined a \textit{diamond} due to its diagrammatically convincing shape. 
In Fig.~\ref{fig:crest-peaks-compare} we show the ratios of absolute complex amplitudes between higher and lower final angular momentum for $q=\pm2$. 
For helium the limit of the ratio is 
\begin{equation}
    \label{eq:two-photon-angular-ratio-he}
    \bigg{|}\frac{\tilde{a}_{\ell_>}^{\pm 2}}{\tilde{a}_{\ell_<}^{\pm 2}}\bigg{|} \to \bigg{|} \frac{\mel{Y_{30}}{Y_{10}}{Y_{20}}\mel{Y_{20}}{Y_{10}}{Y_{10}}}{|\mel{Y_{10}}{Y_{10}}{Y_{20}}|^2 + |\mel{Y_{10}}{Y_{10}}{Y_{00}}|^2}\bigg{|} = \frac{2}{\sqrt{21}},
\end{equation}
while for neon the limit is 
\begin{equation}
    \label{eq:two-photon-angular-ratio-ne}
    \bigg{|}\frac{\tilde{a}_{\ell_>}^{\pm 2}}{\tilde{a}_{\ell_<}^{\pm 2}}\bigg{|} \to \bigg{|}
    \frac{\mel{Y_{41}}{Y_{10}}{Y_{31}}\mel{Y_{31}}{Y_{10}}{Y_{21}}}{|\mel{Y_{21}}{Y_{10}}{Y_{31}}|^2 + |\mel{Y_{21}}{Y_{10}}{Y_{11}}|^2}\bigg{|} = \frac{\sqrt{8}}{3 \sqrt{3}}.
\end{equation}
We note that these ratios are close to one half in both cases, which implies that the lower angular momentum is much favoured over the higher angular momentum in the $q=\pm 2$ peaks.  
Although the matrix elements in the denominator of Eqs.~(\ref{eq:two-photon-angular-ratio-he},\ref{eq:two-photon-angular-ratio-ne}) are taken in absolute-square, one should not misunderstand this is an incoherent summation over paths in the diamond.  
Instead, this indicates that the two coherent paths to the lower angular momentum add up constructively. The fact that the two paths in the diamond add up {\it in phase} with each other can be understood by considering the continuum--continuum phases acquired in laser-stimulated transitions, which only weakly depend on the angular momentum transitions, {\it c.f.} Ref.~\cite{DahlstromJPB2012}. 

In the high-energy limit, the value of the $q=\pm 2$ ratio between final angular momenta $(>/<)$ is explained by a constructive interference effect between different intermediate partial waves, 
while it is the radial integrals that explain the difference between the PADs in absorption $(+)$ and emission $(-)$ at low energies. 
The $q=2$ peak in neon $m=\pm 1$ is a good example of the importance of this interplay between angular and radial integral effects. 
The paths leading to the final lower angular momentum goes through  increasing--decreasing or decreasing--increasing angular momentum pathways in the continuum that are (radially) weaker than the path of two times increasing the orbital angular momentum to the larger final angular momentum. Yet, the constructive interference to the lower angular momentum $\ell=2$, results in a probability ratio strongly favouring the transition that lowers the angular momentum.
In other words: two average paths tend to overtake the one enhanced path. However, in the low energy limit the radial effect can dominate over the interference effect, as evidenced in Fig.~\ref{fig:crest-peaks-compare}~(b) for $q=2$, where the larger angular momentum amplitude is marginally greater than the smaller angular momentum amplitude.

We now turn to the question how the weak radial effects can change the number of minima in the PADs?
The condition of a node in the angular distribution that consists of two partial waves, is found by setting $f(\theta)$ from Eq.~\eqref{eq:fthetavarhpi} to zero. This leads to the following relation:  
\begin{equation}
\label{eq:fiszero}
\frac{\tilde a^{\pm q} _{\ell_>}}{\tilde a^{\pm q}_{\ell_<}} 
= - 
\frac{Y_{\ell_< m}(\theta,\varphi)}{Y_{\ell_> m}(\theta,\varphi)},
\end{equation}
where $\tilde a^{\pm q}_{\ell_{>/<}}$ are different for both absorption $(+)$ and emission $(-)$. 
Interestingly, we have found that the ratios on the right hand side of Eq.~\eqref{eq:fiszero} are equal to 
\begin{align}
    \lim_{\theta\rightarrow\pi/2}-\frac{Y_{00}(\theta,\varphi)}{Y_{20}(\theta,\varphi)}&=\frac{2}{\sqrt{5}}  \\
    \lim_{\theta\rightarrow\pi/2}-\frac{Y_{11}(\theta,\varphi)}{Y_{31}(\theta,\varphi)}&=\sqrt{\frac{8}{7}} \\
    \lim_{\theta\rightarrow\pi/2}-\frac{Y_{10}(\theta,\varphi)}{Y_{30}(\theta,\varphi)}&=\frac{2}{\sqrt{21}}  \\
    \lim_{\theta\rightarrow\pi/2}-\frac{Y_{21}(\theta,\varphi)}{Y_{41}(\theta,\varphi)}&=\frac{\sqrt{8}}{3\sqrt{3}}, 
\end{align}
in the limit of a polar angle equal to 90 degrees, which is equal to the corresponding ratios in Eqs.~(\ref{eq:one-photon-angular-ratio-he}--\ref{eq:two-photon-angular-ratio-ne}). 
This implies that the condition for a node in Eq.~\eqref{eq:fiszero} is {\it just at the limit} for 90 degrees and, therefore, sensitive to small changes in the magnitude of the partial wave amplitudes. 
In the case of absorption, the radial effect allows for nodes at angles close to 90 degrees due to increasing contribution of the higher angular momentum, while in emission this condition will not be satisfied.  
For PADs in helium, shown in Fig.~\ref{fig:raw-XUV38-He1s}~(b,c), this effect explains the two sharp minima on either side of 90 degrees for both $q=1$ and $q=2$. The third sharp minimum at 90 degrees for $q=2$ arises due to the odd parity of the photelectron after absorption of an odd number of photons from the helium ground state. 
In contrast, there is no sharp minima for $q=-1$ and only a single sharp minimum for $q=-2$ due to odd parity in helium. This is because  the condition for additional minima of Eq.~\eqref{eq:fiszero} are not met due to an increased contribution of the lower angular momentum. In $q=-1$ in helium we do observe a local minimum at 90 degrees, which is not fulfilling the condition in Eq.~\eqref{eq:fiszero}.    

For PADs in neon with $m=0$, shown in Fig.~\ref{fig:raw-Ne2p-XUV38-m}~(b,c), the condition for nodes is not satisfied for either $q=\pm 1$.
The $q=1$ case shows two shallow minima, while the $q=-1$ case shows two shoulders, which indicates that emission is further from the additional node condition in Eq.~\eqref{eq:fiszero}. There is a sharp minimum at 90 degrees in both $q=\pm 1$ in neon with $m=0$ due to odd parity after exchange of two photon. For $q=2$ we have three spherical harmomics that interfere. In this case we see are two sharp minima, where the condition of a node is fullfilled, and two additional outer shoulders, where the node condition is not fully satisfied. For $q=-2$ the flat region comes from the fact that the conditions for nodes is not fully met at either of these four instances.  
For neon with $m=\pm 1$, 
shown in Fig.~\ref{fig:raw-Ne2p-XUV38-m}~(e,f), there are again two partial waves that interfere.  
The condition for additional nodes is found in both $q=1$ and $q=2$, while it is not found for $q=-1$ or $q=-2$. The nodes at 0, 90 and 180 degrees are related to the static symmetry properties of spherical harmonics with $m=\pm 1$. 
%


Finally, the absence of a qualitative difference between $q=1$ and $q=-1$ for neon with incoherent addition of both $m=0$ and $m=\pm 1$, can be understood by the fact that $m=0$ is has two small maxima at approximately same angles where $m=\pm 1$ have nodes, as seen in Fig.~\ref{fig:raw-Ne2p-XUV38-m}~(b) and (e), respectively. This effect covers up the difference between absorption and emission processes in the first sideband of neon. This motivates precision experiments with resolution in the magnetic quantum number when studying  laser-assisted photoionization to study the propensity rules in atoms. 
Alternatively, the consequence of additional nodes that come from propensity rule effects can be studied by angle-resolved atomic delay measurements, as shown by Busto~et al.~\cite{busto_fano_2019}. The strong importance of including incoherently both $m=0$ and $m=\pm 1$ contributions for atomic delay simulations in neon was shown by Ivanov and Kheifets~\cite{IvanovPRA2017}. Physically, this is due to the fact that only the absorption paths obtain additional nodes, a criterion formulated in Eq.~\eqref{eq:fiszero}. Each additional node is associated with $\pi$-shifts in absorption paths that leads to strong angle-dependence of atomic delays. Our extension of sideband studies to the second sideband motivates angle-resolved atomic delay experiments with higher-order sidebands, similar to that proposed by Harth et al.~\cite{HarthPRA2019}. 

\section{Conclusions}
\label{sec:conclusions}
With our combined TDCIS and t-SURFF approach, we are able to simulate  angle-resolved photoelectron spectra and identify qualitative differences between sidebands formed by laser-driven absorption and emission processes in the continuum. 
First, we confirm the generalization of Fano's propensity rule to  continuum--continuum absorption and emission processes to the first sideband peaks, then we show that the propensity rule also has  consequences for the second-order sideband peaks. In addition to the propensity rule, we identify that interference effects of different intermediate partial waves plays an important role, which is a stronger effect than the propensity rule at high kinetic energies. 
While Fano's propensity rule for absorption of photons states that an increase of the angular momentum is favoured over a decrease of angular momentum, the interference effect from multiphoton processes can strongly favour a decrease of the angular momentum for both laser absorption and emission processes in the continuum.  
Finally, we find that the propensity rule can be used to explain the  appearance of additional deep minima (nodes) in the angular distributions found in multiphoton absorption processes in the continuum in both the first and second sideband.   

\subsection*{Acknowledgement}
JMD acknowledges support from the Olle Engkvist Foundation, the Knut and Alice Wallenberg Foundation and the Swedish Research Council. 

\appendix
\section{Comparison of the final expression}
\label{apdx:tsurff}
Compared to the expression given in Eq.~(22) by Karamatskou et al.~\cite{Karamatskou_2014}, we do not obtain the same analytical expression.
On the second line of Eq.~\eqref{eq:TSURFF-final}, we have a factor
\begin{equation*}
    \left(kr_c j_{\ell_p}(kr_c) + j_{\ell_p}(kr_c) \right)
\end{equation*}
whereas they have (given in our notation)
\begin{equation*}
    \left( -j_{\ell_p}(k r_c) + \frac{kr_c}{2} j_{\ell_p}^\prime (kr_c) - \frac{1}{2} j_{\ell_p}(kr_c) \right),
\end{equation*}
and on the third line, we have an $r_c$ which is not present in their work.
We speculate that our faster convergence in time may be due to this discrepancy in the analytical expressions.

\bibliographystyle{unsrt}
\bibliography{references}
\end{document}